\documentclass[11pt,a4paper]{article} \usepackage{jheppub}

\title{Constraining BSM Physics at the LHC: Four top
  final states with NLO accuracy in perturbative QCD}

\author[a]{G. Bevilacqua} 
\affiliation[a]{Institut f\"ur Theoretische Teilchenphysik und Kosmologie,
RWTH Aachen University, Otto-Blumenthal Str., D-52056 Aachen, Germany} 
\emailAdd{bevilacqua@physik.rwth-aachen.de} 

\author[b]{and M. Worek} 
\affiliation[b]{Theoretische Physik, Fachbereich C, Bergische 
Universit\"at Wuppertal, Gauss Str. 20, D-42097 Wuppertal, Germany} 
\emailAdd{worek@physik.uni-wuppertal.de} 

\abstract{ Many theories, from Supersymmetry to models of Strong Electroweak
  Symmetry Breaking, look at the production of four top quarks as an
  interesting channel to evidentiate signals of new physics beyond the
  Standard Model. The production of four-top final states requires large
  partonic energies, above the $4m_t$ threshold, that are available at the
  CERN Large Hadron Collider and will become more and more accessible with
  increasing energy and luminosity of the proton beams. A good theoretical
  control on the Standard Model background is a fundamental prerequisite for a
  correct interpretation of the possible signals of new physics that may arise
  in this channel.  In this paper we report on the calculation of the
  next-to-leading order QCD corrections to the Standard Model process $pp\to
  t\bar{t}t\bar{t} + X$. As it is customary for such studies, we present
  results for both integrated and differential cross sections. A judicious
  choice of a dynamical scale allows us to obtain nearly constant ${\cal
    K}$-factors in most distributions.}

\dedicated{\rm WUB/12-12, TTK-12-22}

\begin{document} 
\maketitle
\flushbottom

%

\section{Introduction}

%
Since its discovery in 1995  by the CDF and D0 experiments at
Fermilab, the top quark, the most massive of all the observed
elementary particles, has been extensively scrutinized. Several
properties and observables have been already analysed including, among
others, the top-pair and single-top production cross sections together
with the measurement of the top quark mass. Most of these measurements
have been, however, limited by the relatively small statistics
collected at the Tevatron. With a much higher center of mass energy
and luminosity, the Large Hadron Collider (LHC) has all the features
of a top factory and will significantly add to the previous
measurements in this field. At the nominal design value of the beam
energy ($\sqrt{s} = 14$ TeV) and assuming a luminosity of $10 \mbox{
  fb}^{-1}/\mbox{year}$, the LHC will collect about 8 million top
anti-top pair events and another few million tops via  electroweak
single top quark production. In fact even at present, running at
$\sqrt{s} = 7$ TeV, it has made possible to start re-examining the
main properties of the top quark with significant precision.

The large statistics available will open a window on entirely new
measurements as well as on the analysis of novel, more complex  final
states. As to $t\bar{t}t\bar{t}$ production, the LHC energy is
sufficient to  produce such events at a sensible rate. Assuming only
the standard, dominant $t \to Wb$ decay, this process leads to
$W^+W^-W^+W^-b\bar{b}b\bar{b}$ final states with leptonic and/or
hadronic decays of gauge bosons. The experimental signature of such
final states is characterized by the presence of opposite and
same-sign charged leptons, large missing  energy and jets with a
considerable number of b-jets. These signatures are precisely the ones
that are expected in the Higgs boson(s) and new physics
searches. Thus, $t\bar{t}t\bar{t}$ production is an interesting
channel to probe several realizations of Standard Model extensions at
the LHC. Some prominent examples include models of Higgs and top
compositeness as well as models involving the production of new
colored resonances with large couplings to the top quark. In the
latter case, the most studied ones are Kaluza-Klein gluons from the
Randall-Sudrum warped extra dimensions. Many of these models predict
effective four-top quark interactions as well as new processes such as
$pp\to GG$ and $pp\to t\bar{t} \,G$, where $G$ is a new heavy particle
decaying to $t\bar{t}$, leading to $t\bar{t}t\bar{t}$ final states
(see {\it e.g.} \cite{Dicus:1994sw,Djouadi:2007eg,
  Guchait:2007jd,Lillie:2007hd,Pomarol:2008bh,Kumar:2009vs,Morrissey:2009tf,
  Brooijmans:2010tn,Jung:2010ms,Gregoire:2011ka} and references
therein).  In this context, a precise theoretical description of the
four-top production rate in the Standard Model may help to constrain
new physics scenarios.

In addition, $t\bar{t}t\bar{t}$ is a major background for many
processes arising from supersymmetric extensions of the Standard
Model, the most noticeable example being the production of a heavy
Higgs boson \cite{Gianotti:2002xx,Baur:2002rb}. An
accurate theoretical description of four-top production would help in
this case to determine the Higgs boson self coupling and thus
contribute to better understand the Higgs boson potential in this
minimal SM extension. Furthermore, multi-top final states can also be
produced via long cascade decays of  colored superymmetric particles,
such as squarks or gluinos \cite{Toharia:2005gm,Alves:2011wf}.     

Last but not least, $t\bar{t}t\bar{t}$ production is the last process
in the so-called Les Houches next-to-leading order (NLO)
experimenter's wishlist \cite{Maestre:2012vp} that is still waiting
for a calculation. Recent breakthrough in one-loop calculational
techniques, sometimes referred to as  {\it "the unitary revolution"}
\cite{Bern:1994zx,Britto:2004nc,Ossola:2006us,Giele:2008ve},  together
with great improvement in more traditional methods, have led to a
tremendous progress in the calculation of multi-leg processes at
hadron colliders. This is exemplified by the calculation of the
following $2 \to 4$ processes: $pp(p\bar{p})\to t\bar{t}b\bar{b} + X$
\cite{Bredenstein:2009aj,Bredenstein:2010rs,
  Bevilacqua:2009zn,Worek:2011rd}, $pp(p\bar{p})\to  t\bar{t}jj + X$
\cite{Bevilacqua:2010ve,Bevilacqua:2011aa}, $pp(p\bar{p})\to
W^{+}W^{-}b\bar{b} + X$ \cite{Bevilacqua:2010qb,Denner:2010jp},  $pp
\to b\bar{b}b\bar{b} + X$ \cite{Greiner:2011mp},  $pp(p\bar{p})\to
W^{+}W^{-}jj + X$ \cite{Melia:2011dw,Greiner:2012im}, $pp \to
W^{+}W^{+}jj + X$ \cite{Jager:2009xx,Melia:2010bm},  $pp(p\bar{p})\to
W + 3j + X$ \cite{Berger:2009zg,Berger:2009ep}, $p\bar{p}\to
Z/\gamma^{*} + 3j + X$ \cite{Berger:2010vm},  $pp \to W \gamma\gamma j
+ X$ \cite{Campanario:2011ud} and   $pp \to 4j + X$
\cite{Bern:2011ep}. In addition, the first NLO QCD corrections to
$2\to 5$ processes, {\it i.e.} $pp \to  W + 4j + X$ and $pp \to
Z/\gamma^{*} + 4j + X$  have recently been completed
\cite{Berger:2010zx,Ita:2011wn}.  

In this paper we present the results of a calculation of $pp\to
t\bar{t}t\bar{t} + X$ in the Standard Model at NLO QCD accuracy. This
calculation makes the Les Houches wishlist finally complete, with at
least one independent calculation available for each benchmark
process.  The paper is organized as follows. In Section 2 we briefly
describe the details of our calculation.  Numerical results for the
integrated and differential cross sections are presented in Section
3. Finally, in Section 4 we give  our conclusions.

%

\section{Details of the calculation}

%

\begin{figure}
\begin{center}
\includegraphics[width=0.99\textwidth]{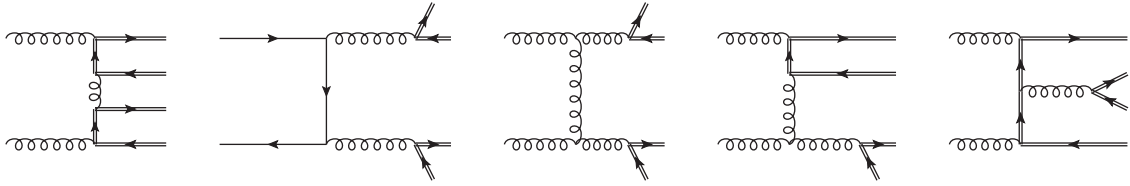}
\end{center}
\caption{\it \label{fig:LO}  A representative set of Feynman diagrams
  contributing to the $pp \to t\bar{t}t\bar{t}$ process at ${\cal
  O}(\alpha_s^4)$. Double lines correspond to top quarks, single
  lines to light quarks and wiggly ones to gluons.}
\end{figure}

At the leading order (LO) in perturbative expansion, $t\bar{t}t\bar{t}$ final
states are produced via the scattering of either two gluons or one quark and
the corresponding anti-quark. A representative set of Feynman diagrams
contributing to the $pp \to t\bar{t}t\bar{t}$ process at ${\cal
  O}(\alpha_s^4)$ is depicted  in Figure \ref{fig:LO}. In total, there are 72
LO  diagrams for $gg \to t\bar{t}t\bar{t}$ and 14 for  $q\bar{q}\to
t\bar{t}t\bar{t}$.  Even though we do not actually employ Feynman diagrams, it
is customary to present them as a measure of the complexity of the
calculation. The calculation of scattering amplitudes is based on well-known
off-shell iterative algorithms
\cite{Draggiotis:1998gr,Draggiotis:2002hm,Papadopoulos:2005ky}, performed
automatically within the \textsc{Helac-Dipoles} package \cite{Czakon:2009ss}
and cross checked with the \textsc{Helac-Phegas} Monte Carlo program
\cite{Kanaki:2000ey,Papadopoulos:2000tt,Cafarella:2007pc}. A perfect agreement
has been found in all cases. Phase-space optimization and integration has been
performed with the help of  \textsc{Parni} \cite{vanHameren:2007pt} and
\textsc{Kaleu} \cite{vanHameren:2010gg}.
%
\begin{figure}
\begin{center}
\includegraphics[width=0.99\textwidth]{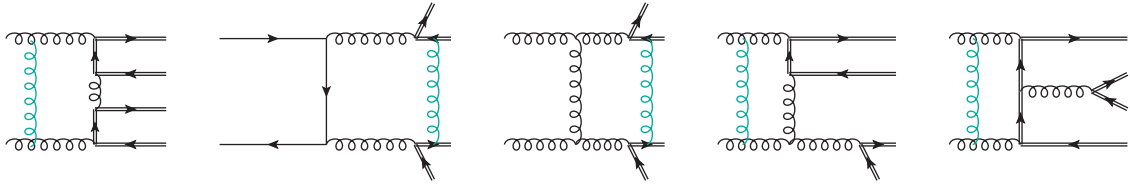}
\end{center}
\caption{\it \label{fig:virtual}   A representative set of pentagon and hexagon
  diagrams for  the $pp \to t\bar{t}t\bar{t} + X$ process at NLO QCD. Double 
  lines correspond to top quarks, single lines to light quarks and wiggly ones 
  to gluons.}
\end{figure}
\begin{figure}
\begin{center}
\includegraphics[width=0.99\textwidth]{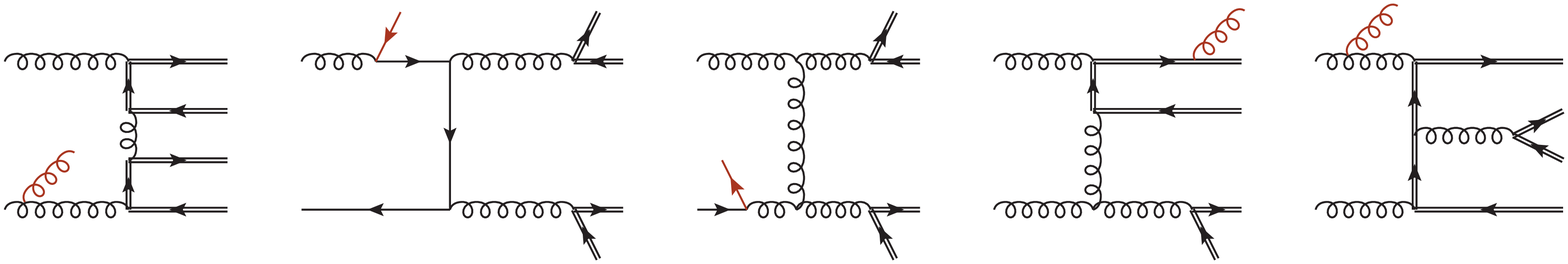}
\end{center}
\caption{\it \label{fig:real}  
A representative set of Feynman diagrams contributing to the real emission 
corrections to the $pp \to t\bar{t}t\bar{t} + X$ process at ${\cal
  O}(\alpha_s^5)$. Double lines correspond to top quarks, single
  lines to light quarks and wiggly ones to gluons.}
\end{figure}
%

At the NLO level, the virtual corrections are obtained from the
interference of the one-loop diagrams with the tree level
amplitude. They can be classified into self-energy, vertex, box-type,
pentagon-type and hexagon-type corrections.  A representative set of
pentagon  and hexagon diagrams is shown in  Figure \ref{fig:virtual}.
The number of one-loop Feynman diagrams for the $pp\to
t\bar{t}t\bar{t}$ process, as obtained with \textsc{Qgraph}
\cite{Nogueira:1991ex}, is 2200 for  $gg\to t\bar{t}t\bar{t}$ and 410
for the $q\bar{q}\to t\bar{t}t\bar{t}$.  Virtual corrections are
evaluated in $d = 4 - 2\epsilon$  dimensions in the  't~Hooft-Veltman
version of the dimensional regularization scheme. The singularities
coming from infrared divergent pieces are canceled by the
corresponding ones arising from the so-called integrated dipoles, {\it
  i.e.} from the counterterms of the adopted subtraction scheme
integrated over the phase space of the unresolved parton. The finite
contributions of the loop diagrams are evaluated numerically in $d =
4$ dimension.  Finally, to ensure numerical stability, we check Ward
identities for every phase space point.  The calculation of the
virtual corrections is achieved with the help of the package
\textsc{Helac-1Loop} \cite{vanHameren:2009dr} which incorporates
\textsc{CutTools} \cite{Ossola:2007ax,Draggiotis:2009yb} and
\textsc{OneLOop} \cite{vanHameren:2010cp} as cornerstones. The first
code contains an implementation of the OPP method for the reduction of
one-loop amplitudes at the integrand level, while the second one is
dedicated to the evaluation of the one-loop scalar
functions. Renormalization is done, as usual, by evaluating tree-level
diagrams with counterterms.  For our process, we choose to renormalize
the coupling in the $\overline{\rm {MS}}$ scheme with five active
flavors and the top quark decoupled. The mass renormalization is
performed in the on-shell scheme.

The real emission corrections to the LO process arise from tree-level
amplitudes with one additional parton, {\it i.e.} an additional gluon,
or a quark anti-quark pair replacing a gluon. All possible
contributions can be classified into the four categories presented in
Table \ref{tab:real}, together with the number of Feynman diagrams and
the Catani-Seymour dipoles corresponding to each subprocess. Typical
examples of the real emission graphs are displayed in Figure
\ref{fig:real}.  

For the calculation of the real emission contributions, the package
\textsc{Helac-Dipoles} \cite{Czakon:2009ss} is employed. It implements
the massless dipole formalism of  Catani and Seymour
\cite{Catani:1996vz},  as well as its massive version as developed by
Catani, Dittmaier, Seymour and Trocsanyi \cite{Catani:2002hc}, for
arbitrary  helicity eigenstates of the external partons. A phase space
restriction on the  contribution of the dipoles as introduced by Nagy
and Trocsanyi \cite{Nagy:1998bb} is also included. For the real
corrections as well, we adopt \textsc{Kaleu} equipped with additional,
special dipoles channels that proved to be important for  phase-space
optimization in the subtracted real emission part
\cite{Bevilacqua:2010qb}.

To summarize, our computational system relies on \textsc{Helac-1Loop}
and \textsc{Helac-Dipoles} that are both parts of the
\textsc{Helac-NLO} framework for NLO QCD calculations
\cite{Bevilacqua:2011xh}.  The framework is publicly
available\footnote{ \tt
  http://helac-phegas.web.cern.ch/helac-phegas/}.  All numerical
results are obtained using the same methods as presented in our
previous work, therefore we will not describe them here in more
detail. Rather, we address the interested reader to our earlier
publications in this field
\cite{Bevilacqua:2009zn,Worek:2011rd,Bevilacqua:2010ve,
  Bevilacqua:2011aa,Bevilacqua:2010qb}. We would like to emphasize,
however,  that all parts of the calculation are performed in a
completely automatic  and fully numerical way. 

%
\begin{table}
\begin{center}
  \begin{tabular}{||c|c|c||}
\hline\hline
   \textsc{Partonic } & \textsc{Number Of Feynman}  & \textsc{Number Of}\\
   \textsc{Subprocess} &  \textsc{Diagrams}  & \textsc{Dipoles}\\
\hline
$gg \to  t\bar{t}t\bar{t} g$ &  682 & 30 \\ 
$q\bar{q} \to  t\bar{t}  t\bar{t}g $ & 128 & 30 \\
$gq \to   t\bar{t}  t\bar{t} q $ & 128 &10\\
$qg  \to  t\bar{t}  t\bar{t} q $ & 128 & 10 \\
\hline\hline
  \end{tabular}
\end{center}
  \caption{\it \label{tab:real} Partonic subprocesses contributing to
    the subtracted real emissions  at ${\cal{O}}(\alpha_{s}^5)$  for
    the $pp\to t\bar{t}t\bar{t} +X$ process. Also, the number of Feynman
    diagrams and  the number of Catani-Seymour dipoles corresponding to these
    subprocesses are presented. }
\end{table}
%


\section{Predictions for the LHC}

%
We consider the process $pp \to t\bar{t}t \bar{t} +X$ at the LHC with
a nominal design center of  mass energy of $\sqrt{s} = 14$ TeV. For
the top quark mass, we use the Tevatron average value $m_t$ = 173.2
GeV as measured by  the CDF and D0 experiments.   Our calculation,
like any fixed-order one, embodies a residual dependence  on the
renormalization scale ($\mu_R$) and the factorization scale ($\mu_F$)
arising from truncation of the perturbative expansion. As a
consequence, the value of observables depends on the values of $\mu_R$
and $\mu_F$ that are provided as input parameters. For many processes,
a {\it natural scale} can be easily identified in several ways. For
example, one may consider the mass of the heavy particle appearing in
the process, or even the typical momentum transfer or the total
transverse energy of the process. In the case of $t\bar{t}t\bar{t}$,
setting $\mu_0=\mu_R=\mu_F=2m_t$ seems a quite natural scale
choice. In fact, when considering total cross sections, effects of the
phase space regions close to the top production threshold are expected
to dominate, which justifies our choice. 

We consider the MSTW2008 set of parton distribution functions
\cite{Martin:2009iq} as our default PDF set. In particular, we take
MSTW2008lo68cl PDFs with 1-loop running $\alpha_s$ at LO and
MSTW2008nlo68cl PDFs with 2-loop running $\alpha_s$ at NLO,  including
five active flavors.  The strong coupling constant is provided by  the
PDF set itself.  It should be noticed that the contribution induced by
bottom-quark densities amounts to $0.05\%$ at the LO. We decided
therefore to neglect the contribution of the $b\bar{b}$ initial state
in the computation of the total cross section at both LO and NLO.
Outgoing top and anti-top quarks are treated on-shell without any cut
restriction on them.  At NLO, an additional final-state parton
arises. Since this parton has no other partons in the final state to
be recombined with (the tops are always assumed to be tagged), it will
be the only responsible for the possible additional jet. As a
consequence, the properties of the resulting jet are independent of
the choice of  the jet algorithm and the size of the cone radius. We
set no restriction on the kinematics of the extra jet.

%

\subsection{Integrated cross section and its scale dependence for 
$\mu_0= 2m_t$}

%
\begin{table}
\begin{center}
\vspace{0.4cm}
  \begin{tabular}{||c|c|c|c|c|c||}
\hline
\hline
 \textsc{Process} &  $\sigma_{\rm LO}$ [fb]
  & $\sigma_{\rm NLO}^{\rm \alpha_{max}=1}$ [fb] 
  & $\sigma_{\rm NLO}^{\rm \alpha_{max}=0.01}$ [fb] 
  & \textsc{K-Factor}
  & $[\%]$
    \\
\hline\hline
$pp \to t \bar{t} t \bar{t} + X$ & 12.056(6) & 15.33(2) & 15.35(3) 
& 1.27 & 27\\
\hline
\hline
  \end{tabular}
\end{center}
  \caption{\it \label{tab:1} Integrated cross section at LO and NLO
    for $pp \to t \bar{t} t \bar{t} + X $ production at the LHC with
    $\sqrt{s} = 14$ TeV. Results for the  MSTW2008 PDF set are
    presented. In the last two columns  the K factor, defined as the
    ratio of the NLO cross section to the respective LO result, and
    NLO corrections in $\%$ are given. The scale choice is $\mu_F =
    \mu_R = \mu_0 =2 m_t$.}
\end{table}
\begin{figure}
\begin{center}
\includegraphics[width=0.49\textwidth]{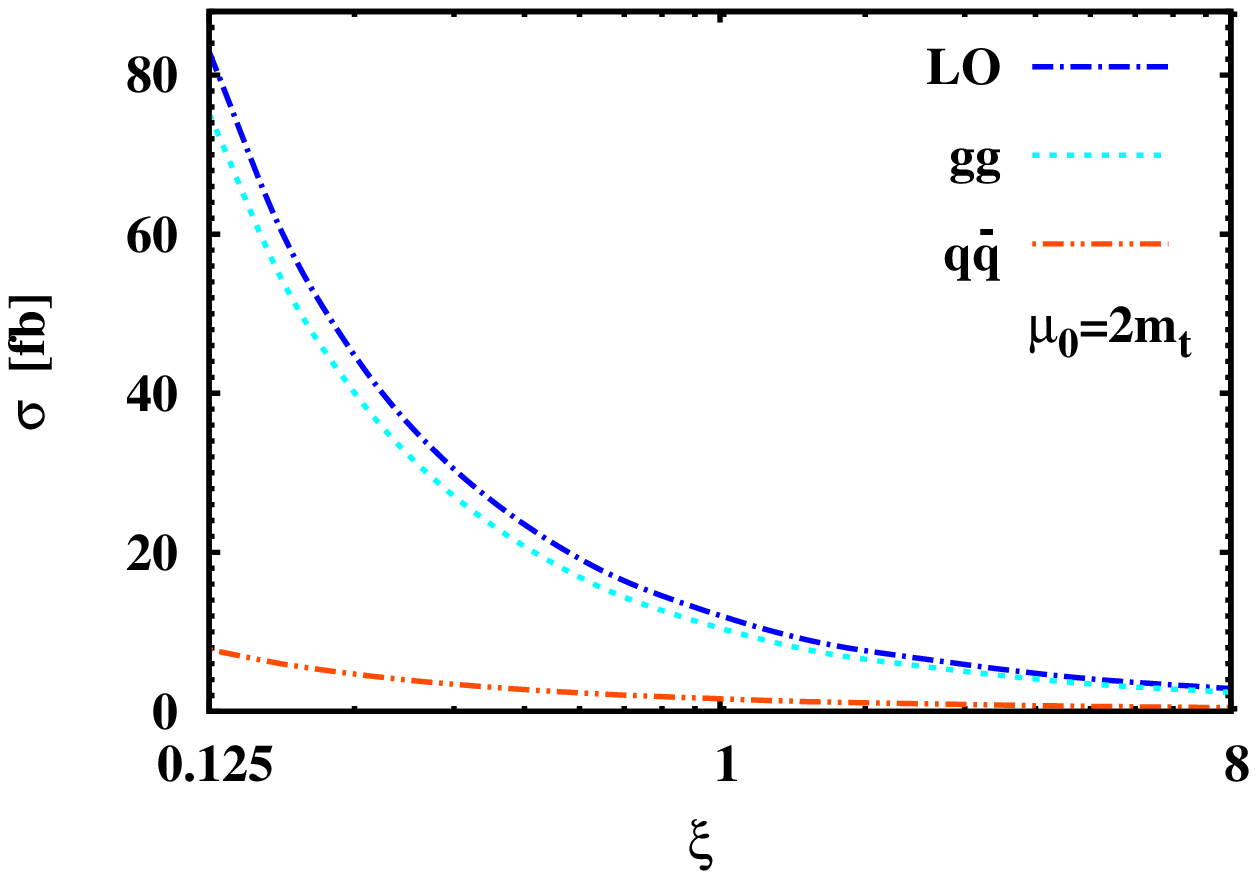}
\includegraphics[width=0.49\textwidth]{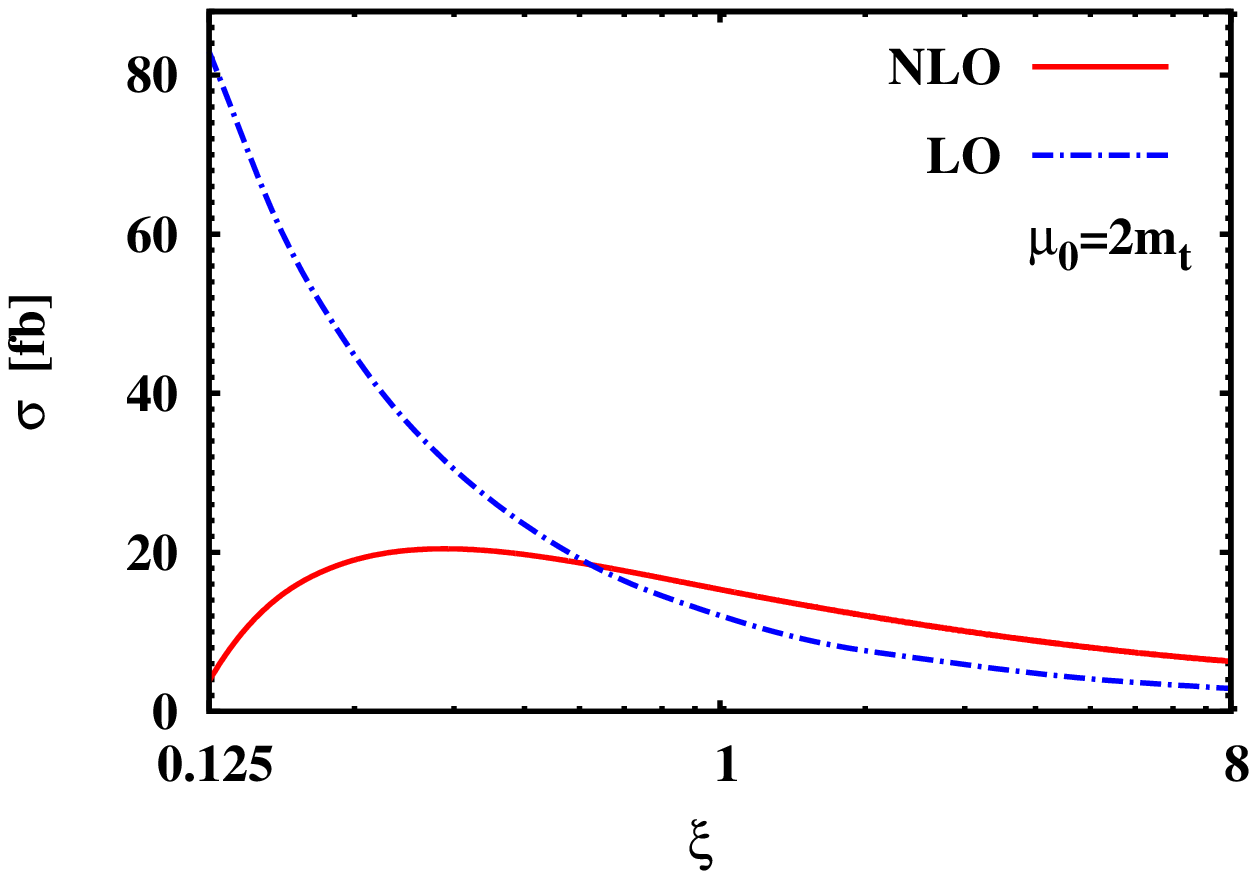}
\end{center}
\caption{\it \label{fig:scale-fixed} Scale dependence of the LO cross
  section with the individual contributions of the partonic channels
  (left panel) and  scale dependence of the LO and NLO cross sections
  (right panel)  for the  $pp\rightarrow t\bar{t} t\bar{t} ~+ X$
  process at the LHC for  $\sqrt{s}=14$ TeV. The scale is set to a
  common value $\mu_R=\mu_F= \xi \cdot \mu_0$ where  $\mu_0=
  2m_t$.}
\end{figure}
%
We start our presentation of the results with the integrated cross sections. In
Table  \ref{tab:1}  cross sections at LO and NLO for $pp \to
t\bar{t}t\bar{t}  + X$ production at the LHC with $\sqrt{s} = 14$ TeV,
are presented.  Our results  for  $\mu_0
= 2m_t$ scale using the MSTW2008 PDF set are

\begin{equation}
\sigma^{\rm{LO}}_{t\bar{t}t\bar{t}}({\rm LHC}_{14 \rm{TeV}}, m_t=173.2 ~{\rm GeV}, 
{\rm MSTW2008lo}) 
= 12.056 ^{+9.364(+78\%)}_{-4.876(-40\%)}~{\rm fb}
\end{equation}
\begin{equation}
\sigma^{\rm{NLO}}_{t\bar{t}t\bar{t}}({\rm LHC}_{14 \rm{TeV}}, m_t=173.2 ~{\rm GeV},
 {\rm MSTW2008nlo}) 
= 15.33 ^{+3.95(+26\%)}_{-3.81(-25\%)} ~{\rm fb .}
\end{equation}

This leaves us with an NLO ${\cal K}-$factor equal to ${\cal K} =
1.27$ and a positive NLO QCD  correction of the order of $27\%$.  Also
given in Table \ref{tab:1} are the integrated NLO cross sections for
two values of the unphysical cutoff parameter $\alpha_{\rm max}$ that
is a common modification of subtraction terms in the phase space away
from the singularity, first introduced in \cite{Nagy:1998bb}. To be
more specific we use two values, $\alpha_{\rm max} = 1$ that
corresponds to the original formulation of
\cite{Catani:1996vz,Catani:2002hc},  and $\alpha_{\rm max} =
0.01$. The independence of the final result on the value of the
$\alpha_{\rm max}$ parameter is  clearly visible in the Table
\ref{tab:1}.  This is a strong consistency check of the calculation of
the real emission part. For more details on the $\alpha_{\rm max}$
implementation in the \textsc{Helac-Dipoles} package see
{\it e.g.} \cite{Czakon:2009ss,Bevilacqua:2009zn}. 

The theoretical uncertainty of the total cross section, associated
with neglected higher order terms in the perturbative expansion, can
be estimated by varying the renormalization and factorization scales
in $\alpha_s$  and PDFs, up and down by a factor 2 around the central
scale of the process, {\it i.e.} $\mu_0$. An observed change  in the value
of  $\sigma^{\rm LO}_{t\bar{t}t\bar{t}}$  for this two scale choices
{\it i.e.}  $0.5\mu_0$ and $2\mu_0$ is truly asymmetric. Taken very
conservatively,  as a maximum  of these two results, the scale
uncertainty at LO is estimated to be  at the level of $78\%$.
However, in case like this  it is more appropriate to symmetrize the
errors.   After symmetrization the scale uncertainty at LO is
assessed to be instead of the order of $59\%$.  After inclusion of the
NLO QCD corrections, the scale uncertainty  is  reduced down to $26
\%$.  In Figure \ref{fig:scale-fixed} a graphical presentation of the
scale dependence is given, both at the LO and NLO. We observe a
dramatic reduction of the scale uncertainty while going from LO to
NLO.  Figure \ref{fig:scale-fixed} also shows the scale dependence of
the LO cross section with the individual contributions of the two
partonic channels. At the central scale, $\mu_0=2m_t$, the $gg$
channel dominates the total $pp$ cross section by about $87\%$
followed by the $q\bar{q}$ channel with about $13\%$. 

The theoretical uncertainty as obtained from the scale dependence of the
cross section is not the only source of systematic  uncertainties.
Another  source of uncertainties comes from the parameterization of
PDFs. We  estimate PDF uncertainties via the so called Hessian method
where, besides the best fit PDFs, a set of 40 PDF parameterizations is
provided that describes $\pm 1 \sigma$ variation of all parameters
which have been used to obtain the global fit. These uncertainties are
due to experimental errors in the various data that are used in the
fits.  We adopt  the prescription from Ref.
\cite{Nadolsky:2001yg,Cacciari:2008zb} and based on these 40 sets
calculate asymmetric uncertainties that estimate maximal variations of
the result calculated for the central value in the positive and
negative directions.  They amount to $+5.7\%$ and $-4.5\%$. However,
this method does not account for the theoretical assumptions that
enter into parameterization of the PDFs which are difficult to
quantify within a given scheme.  Therefore, in the next step we use
a different PDF set, namely CTEQ PDF \cite{Lai:2009ne,Lai:2010vv} that
should have different theoretical assumptions. Specifically, we employ
CT09MC1 PDFs with 1-loop running $\alpha_s$ at LO and CT10 PDFs with
2-loop running $\alpha_s$ at NLO.  We compare the results for the
central value using the best fit PDFs. With $\mu_0=2m_t$  our findings
can be summarized as follows:
\begin{equation}
\sigma^{\rm{LO}}_{t\bar{t}t\bar{t}}({\rm LHC}_{14 \rm{TeV}}, m_t=173.2 ~{\rm GeV}, 
 {\rm CT09MC1}) 
=  11.414(8)  ~{\rm fb}\,, 
\end{equation}
\begin{equation}
\sigma^{\rm{NLO}}_{t\bar{t}t\bar{t}}({\rm LHC}_{14 \rm{TeV}}, m_t=173.2 ~{\rm GeV},
 {\rm CT10}) 
= 14.37(2)   ~{\rm fb} \,.
\end{equation}
The MSTW2008 results are larger than the CTEQ predictions by $5.6\%$
at LO and $6.7\%$ at NLO, which is comparable to the individual
estimates of MSTW2008 PDF systematics\footnote{Let us point out here
  that for the old CTEQ6 PDF sets \cite{Pumplin:2002vw,Stump:2003yu}
  that are still widely used in the phenomenological studies at the
  LHC we have obtained  the following results $\sigma^{\rm
    LO}_{t\bar{t}t\bar{t}}({\rm LHC}_{14 \rm{TeV}}, m_t=173.2 ~{\rm
    GeV},  {\rm CTEQ6L1})  = 8.259(4)$ fb, $\sigma^{\rm
    NLO}_{t\bar{t}t\bar{t}}({\rm LHC}_{14 \rm{TeV}}, m_t=173.2 ~{\rm
    GeV}, {\rm CTEQ6M})  =  14.74(2)$ fb.}.   In fact, since different
values of $\alpha_s$ are associated with the MSTW2008 and the CTEQ
sets one should also consider the uncertainties coming from the
determination of the value of the strong coupling constant,  which is
fitted together with the PDFs. Clearly, further studies are needed to
clarify these issues. However, since the PDF uncertainties for the
process under scrutiny are well below the theoretical uncertainties
due to scale dependence, which remain the dominant source of the
theoretical systematics, we leave such studies for the future.
%
\begin{figure*}
\begin{center}
\includegraphics[width=0.49\textwidth]{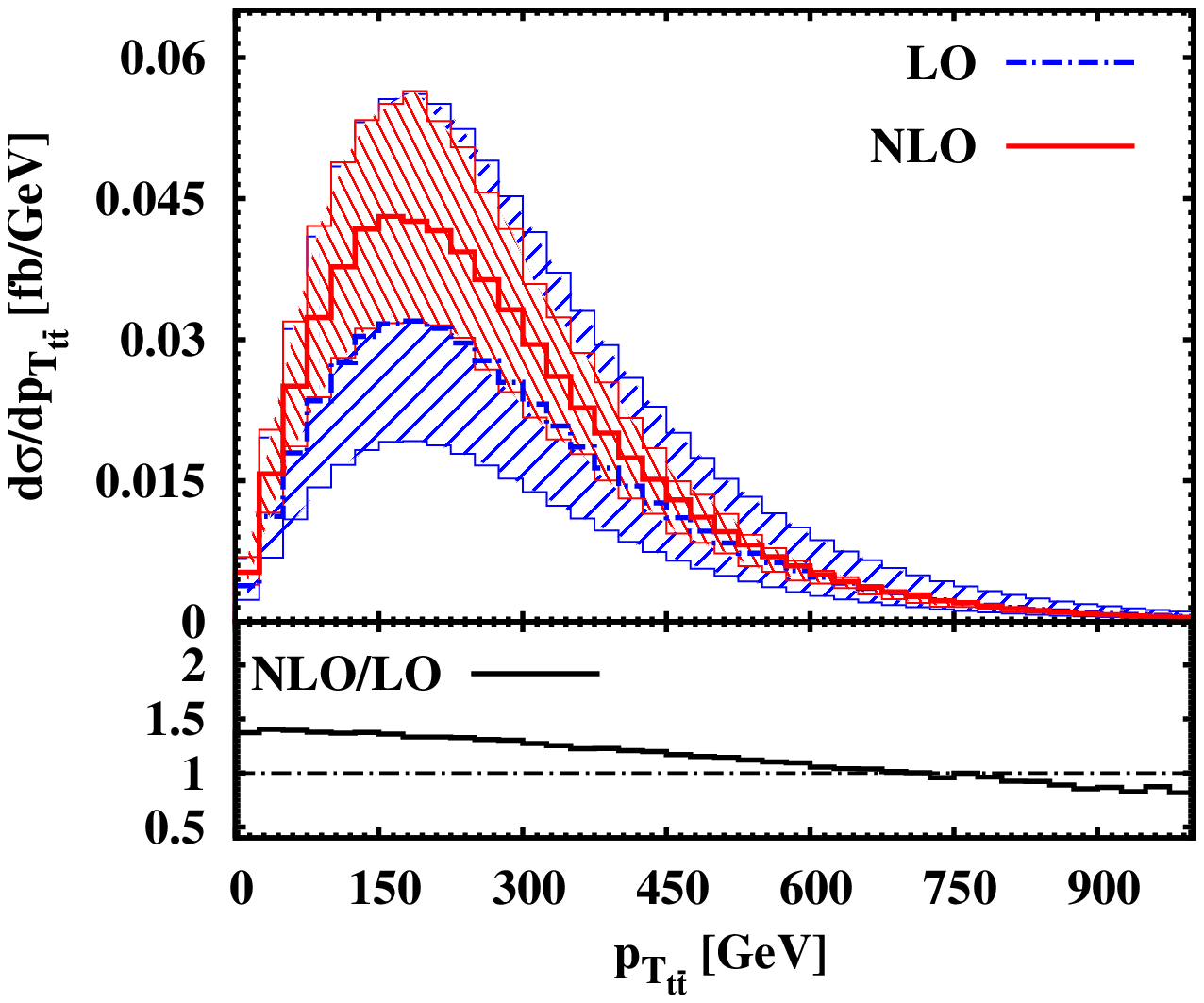}
\includegraphics[width=0.49\textwidth]{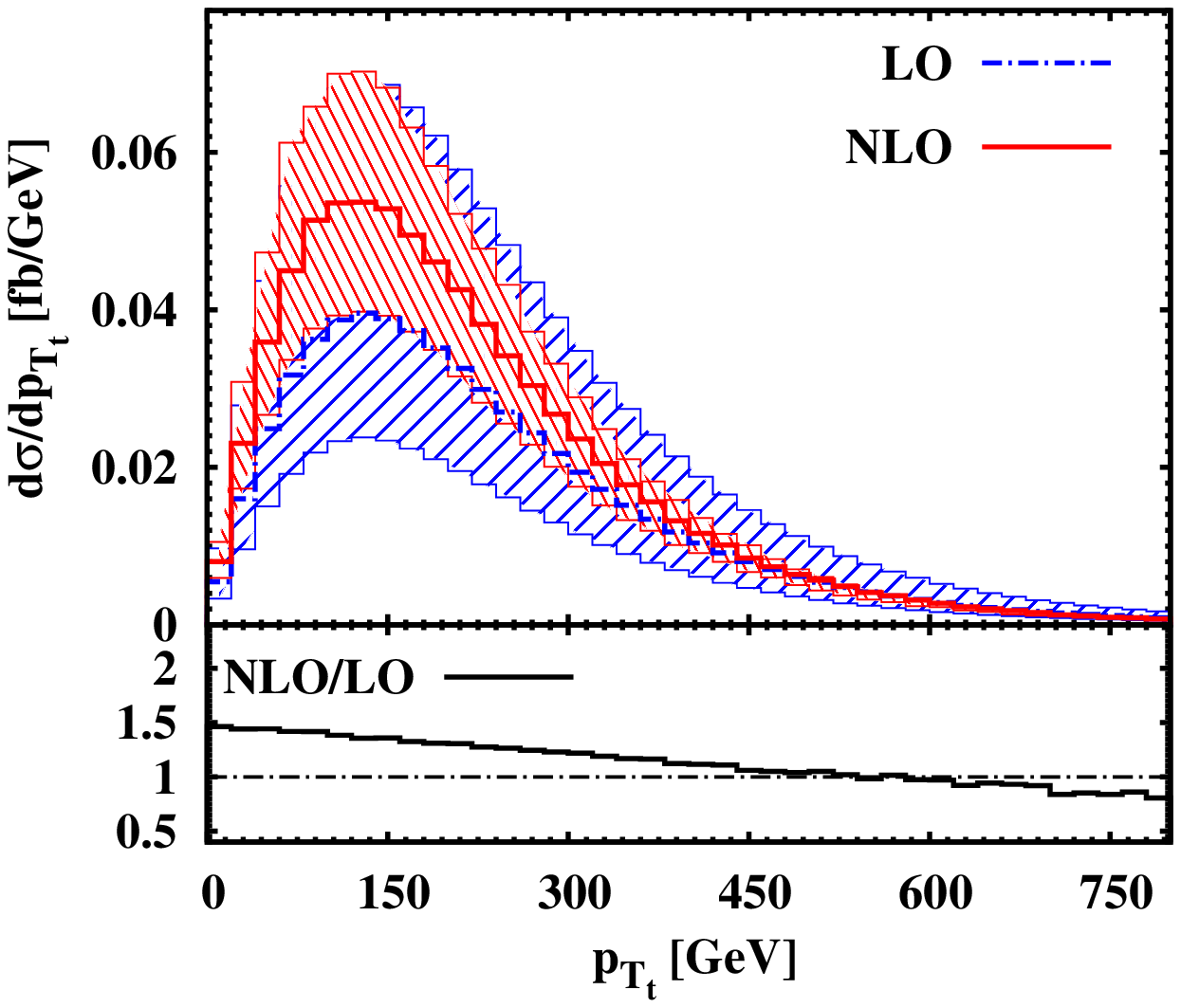}
\includegraphics[width=0.49\textwidth]{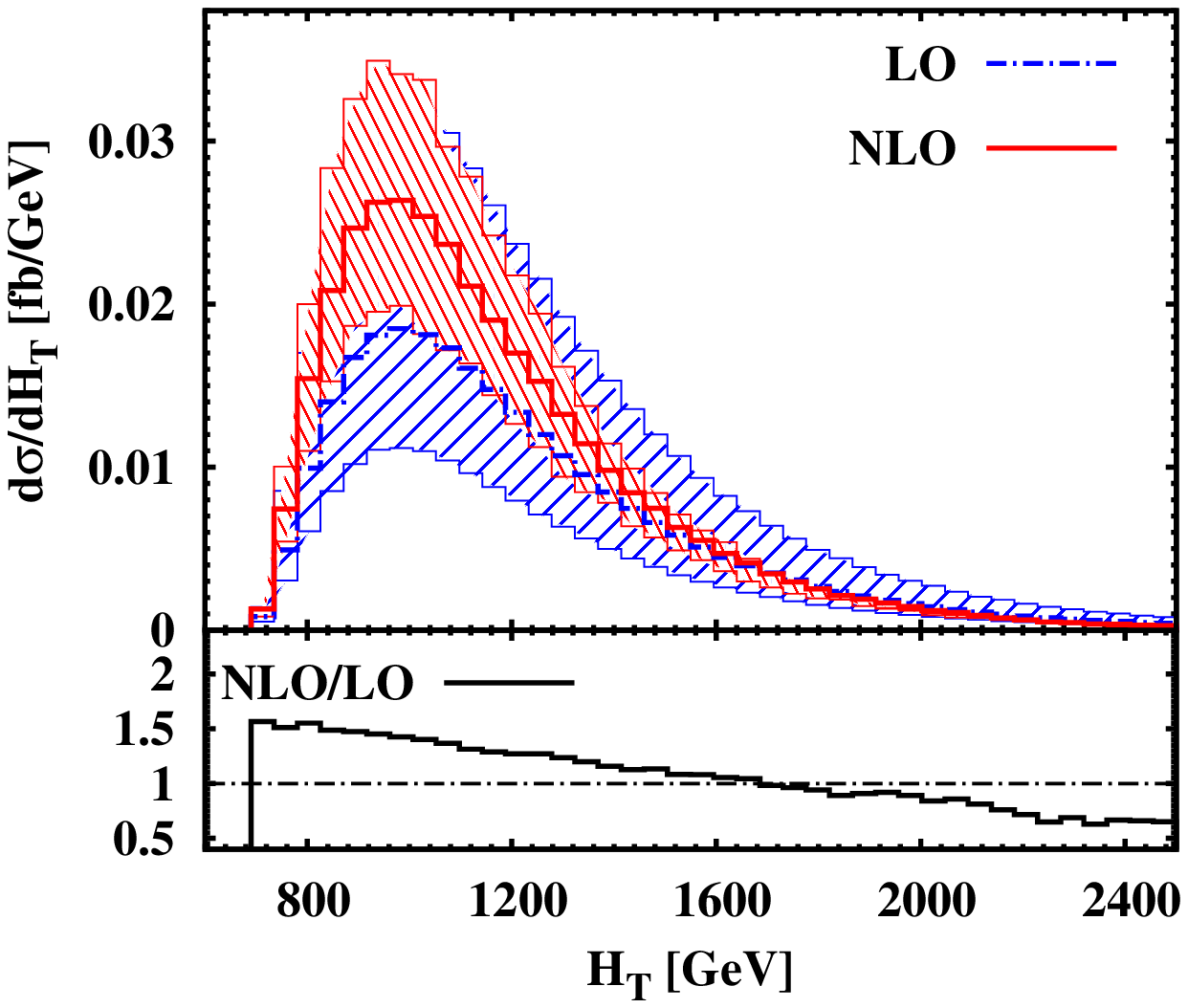}
\end{center}
\caption{\it  \label{lhc:distributions-fixed}   Averaged differential cross
  section distributions as a function of  transverse momentum of the
  $t\bar{t}$ pair (upper-left panel)  and the top quark (upper-right panel)
  for  $ pp \to t \bar{t}  t \bar{t} + X$ production at the LHC with
  $\sqrt{s}= 14 ~\textnormal{TeV}$.  Also shown is the differential  cross
  section distribution as a function of the total transverse energy of the
  system (lower panel).  The dash-dotted (blue) curve corresponds to the LO,
  whereas the solid (red) one to the NLO result. The scale choice is $\mu_F =
  \mu_R =\mu_0= 2 m_t$. The uncertainty bands depict
  scale variation. The lower panels display the differential $\cal K$
  factor.}
\end{figure*}
%

\subsection{Differential cross sections for $\mu_0 = 2m_t$}

%
As already mentioned, total cross sections  are mostly influenced by
final-state production relatively close to the threshold as defined by
particle masses. On the other hand, differential cross sections extend
themselves up to energy scales that are much larger than the
threshold, and may show larger shape distortions in such high-energy
regions. Therefore, in the next step we turn our attention to the
differential  cross sections. 

We have checked as many as 16 observables. Here we present three cases
where the differential ${\cal K}$ factor, defined as the bin-by-bin
ratio of the NLO result to the LO one for the central scale value
$\mu_0$, has been found to be mostly distorted.  In Figure
\ref{lhc:distributions-fixed}   we present  the averaged transverse
momentum distribution of $t\bar{t}$ pair and of the top quark together
with the distribution  of the total transverse energy of the
$t\bar{t}t\bar{t}$ system. The dash-dotted (blue) curve corresponds to
the LO, whereas the solid (red) one to the NLO result. The upper
panels show the distributions themselves and additionally include the
scale-dependence bands  obtained with a  variation of the central
scale  by a factor of two. The lower panels display  the differential
${\cal K}$ factor. 

We observe that at the LHC, employing a fixed scale $\mu_0 = 2m_t$,
the NLO corrections to the transverse momentum distributions do not
simply rescale the LO  shapes, but also induce distortions at the level
of $60\%$. In case of the  total transverse energy of the system,
which we define  as  a sum of (anti-)top quark transverse energies 
\begin{equation}
 H_T= \sum_{i=1,2} E_{T}(t_i) +  \sum_{i=1,2} E_{T}(\bar{t}_i)  \,,  
\end{equation} 
\begin{equation}
 E_{T}(t)= \sqrt{m_t^2 + p_{T}^2(t)}\, ,
\end{equation} 
the situation is even more severe with an observed distortion at the
level of $80\%$ or more.  Clearly, large and negative NLO corrections
affect the high $H_T$ region.  One can also note that the NLO error
bands, estimated through scale variation, do not fit nicely within the
LO ones at low $p_T$'s, as one should expect from a well-behaved
perturbative expansion.  We can summarize our conclusions at this
stage by remarking that the fixed-scale choice
$\mu_R=\mu_F=\mu_0=2m_t$ does not ensure stable shapes when going from
LO to NLO. Therefore, differential cross sections are properly
described only when the NLO QCD corrections are taken into account.

%

\subsection{Integrated cross section and its scale dependence 
for $\mu_0 = H_T/4$}

%
With the goal of stabilizing shapes in the high $p_T$ and $H_T$
regions, that are relevant for the new physics searches, we have
explored a dynamical choice for $\mu_R$ and $\mu_F$ that helps to
achieve flatter differential ${\cal K}$-factors, thus describing more
appropriately the kinematics of the process far away from the
threshold. To this end, we have incorporated the dynamical scale
option into the \textsc{Helac-Nlo} framework. The implementation has
been carefully cross-checked by  reproducing partial and total cross
section results for the NLO QCD corrections to the $pp \to
t\bar{t}b\bar{b}+X$ process with a dynamical scale  for the setup number
I as given in Ref. \cite{Bredenstein:2010rs}. A per-mille level agreement
has been found in all cases.

For the process at hand, we explored several possibilities and decided
in the end to consider the dynamical scale
$\mu_R=\mu_F=\mu_0=H_T/4$\footnote{The sum of  the transverse
  energies of massless outgoing partons and leptons, has already been
  advocated as a good scale choice  in the study of NLO QCD
  corrections to the differential distributions for $pp  \to V + 3j$
  process, where $V=W^{\pm},Z/\gamma^{\star}$
  \cite{Berger:2009ep,Berger:2010vm}.}.  While preserving moderate NLO
QCD corrections and the dramatic reduction of the theoretical
uncertainty, this new scale choice turns out to be particularly
effective in improving the stability of distribution shapes.
%
\begin{figure}
\begin{center}
\includegraphics[width=0.49\textwidth]{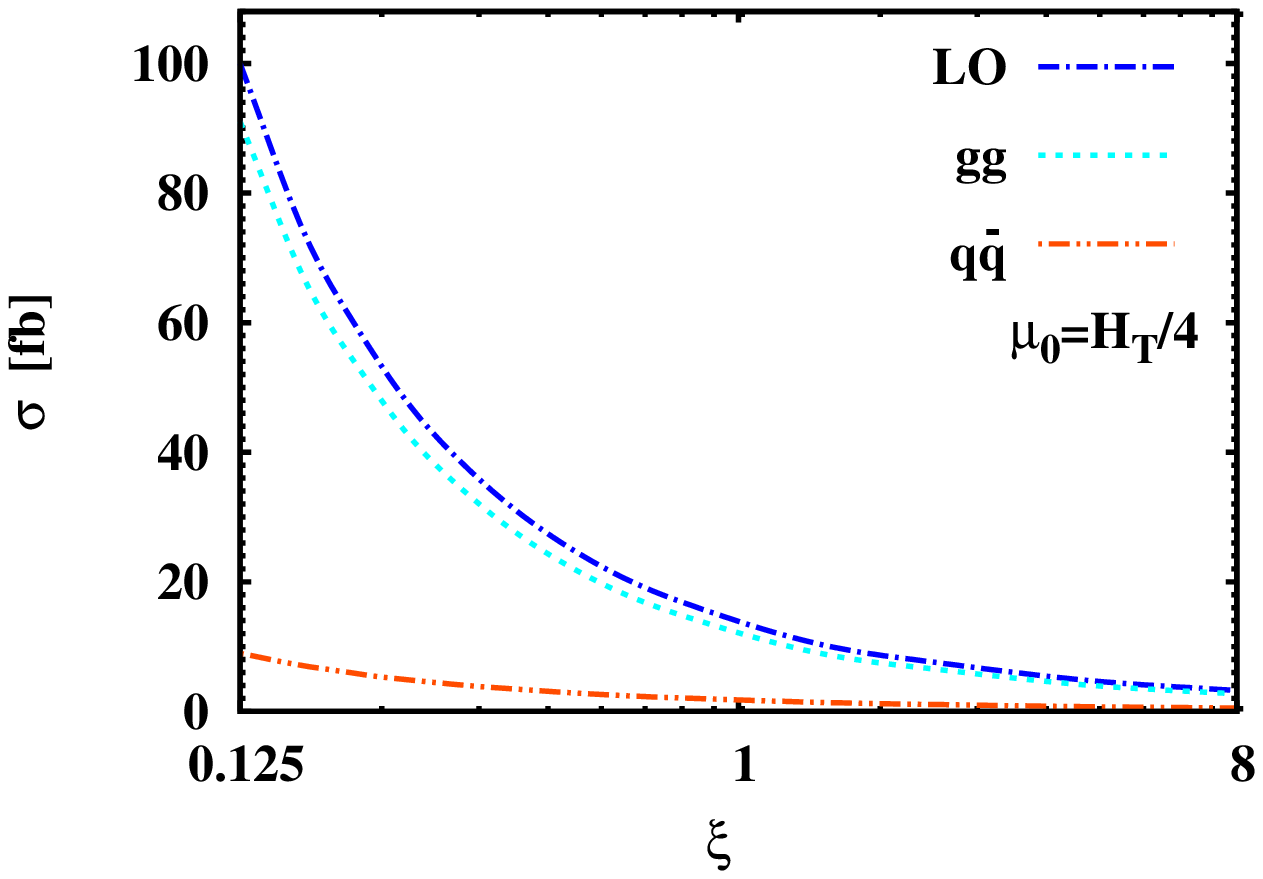}
\includegraphics[width=0.49\textwidth]{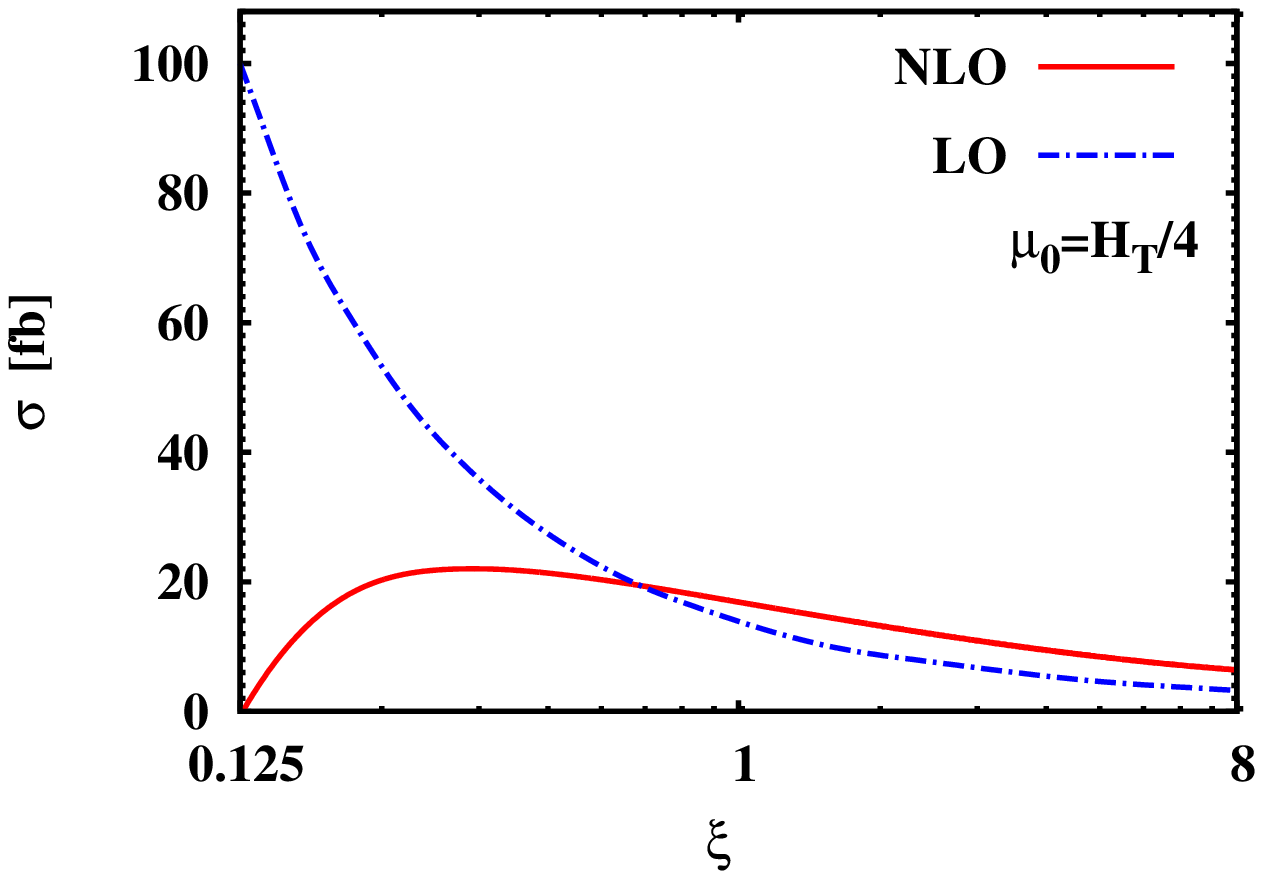}
\end{center}
\caption{\it \label{fig:scale-dynamic} Scale dependence of the LO cross
  section with the individual contributions of the partonic channels
  (left panel) and  scale dependence of the LO and NLO cross sections
  (right panel)  for the  $pp\rightarrow t\bar{t} t\bar{t} ~+ X$
  process at the LHC for  $\sqrt{s}=14$ TeV. The scale is set to a
  common value $\mu_R=\mu_F= \xi \cdot \mu_0$ where  $\mu_0=
  H_T/4$.}
\end{figure}
\begin{table}
\begin{center}
\vspace{0.4cm}
  \begin{tabular}{||c|c|c|c|c|c||}
\hline
\hline
 \textsc{Process} &  $\sigma_{\rm LO}$ [fb]
  & $\sigma_{\rm NLO}^{\rm \alpha_{max}=1}$ [fb] 
  & $\sigma_{\rm NLO}^{\rm \alpha_{max}=0.01}$ [fb] 
  & \textsc{K-Factor}
  & $[\%]$
    \\
\hline\hline
$pp \to t \bar{t} t \bar{t} + X$ & 13.891(9)  & 16.87(2) &  16.86(3)
&  1.21 & 21 \\
\hline
\hline
  \end{tabular}
\end{center}
  \caption{\it \label{tab:2} Integrated cross section at LO and NLO
    for $pp \to t \bar{t} t \bar{t} + X $ production at the LHC with
    $\sqrt{s} = 14$ TeV. Results for the  MSTW2008 PDF set are
    presented. In the last two columns  the K factor, defined as the
    ratio of the NLO cross section to the respective LO result, and
    NLO corrections in $\%$ are given. The scale choice is $\mu_F =
    \mu_R = \mu_0 = H_T/4$.}
\end{table}
%
 Table \ref{tab:2} shows the integrated cross sections at LO and NLO
 for $pp \to t\bar{t}t\bar{t}  + X$ production at the LHC with
 $\sqrt{s} = 14$ TeV, using the same   settings as before with the
 only exception of the scale choice, this time set to be $\mu_0 =
 H_T/4$.  Our results can be summarized as follows:
\begin{equation}
\sigma^{\rm{LO}}_{t\bar{t}t\bar{t}}({\rm LHC}_{14 \rm{TeV}}, m_t=173.2
~{\rm GeV},  {\rm MSTW2008lo})  = 13.891
^{+11.074(+80\%)}_{~-5.711(-41\%)} ~{\rm fb}\, ,
\end{equation}
\begin{equation}
\sigma^{\rm{NLO}}_{t\bar{t}t\bar{t}}({\rm LHC}_{14 \rm{TeV}},
m_t=173.2 ~{\rm GeV}, {\rm MSTW2008nlo})  = 16.87
^{+4.04(+24\%)}_{-4.26(-25\%)} ~{\rm fb} \,.
\end{equation}
The new results are a bit higher, {\it i.e.} by $15\%$ at LO and by $10\%$
at NLO compared with Table \ref{tab:1}, which is perfectly within
theoretical error estimates at the corresponding perturbative order
level. Moreover, the ${\cal K-}$factor obtained with this new scale is
smaller, of the order of  ${\cal K} = 1.21$.  The independence of the
final result on the value of the $\alpha_{\rm max}$ parameter has also
been checked in this case as shown in Table \ref{tab:2}.  For
completeness, we show in Figure \ref{fig:scale-dynamic} the scale
dependence of the LO and NLO  cross sections. Again we observe a
striking reduction of the scale uncertainty  while going from LO to
NLO. Varying the scale up and down by a factor 2 changes the cross
section by $+80\%$ and $-41\%$ in the LO case, whereas in the NLO case
we obtain a variation of $+24\%$ and $-25\%$.  With the evaluation of
NLO QCD corrections the theoretical error has been decreased from about
$80\%$ $(60\%)$ down to $25\%$. 
%
\begin{figure*}
\begin{center}
\includegraphics[width=0.49\textwidth]{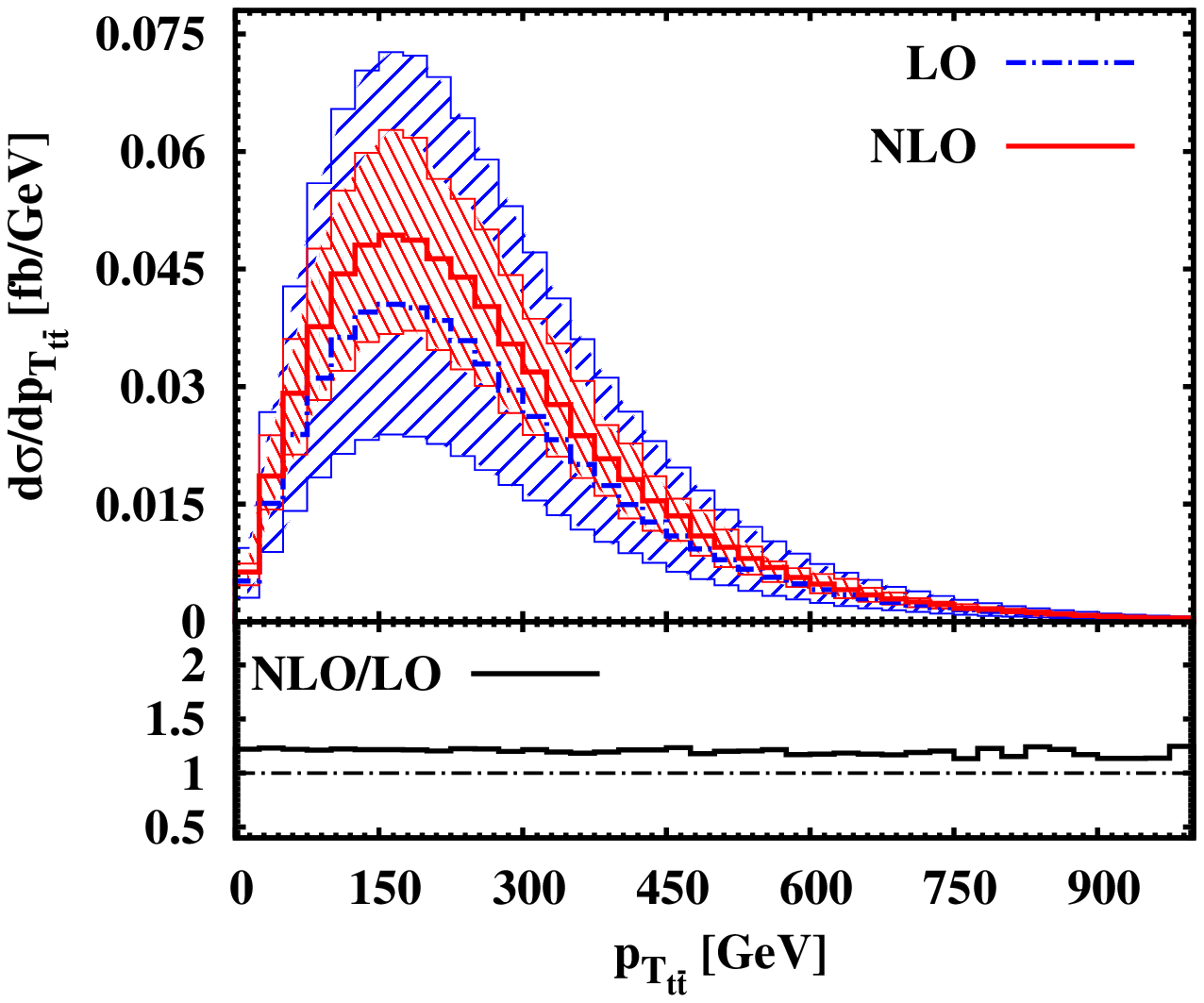}
\includegraphics[width=0.49\textwidth]{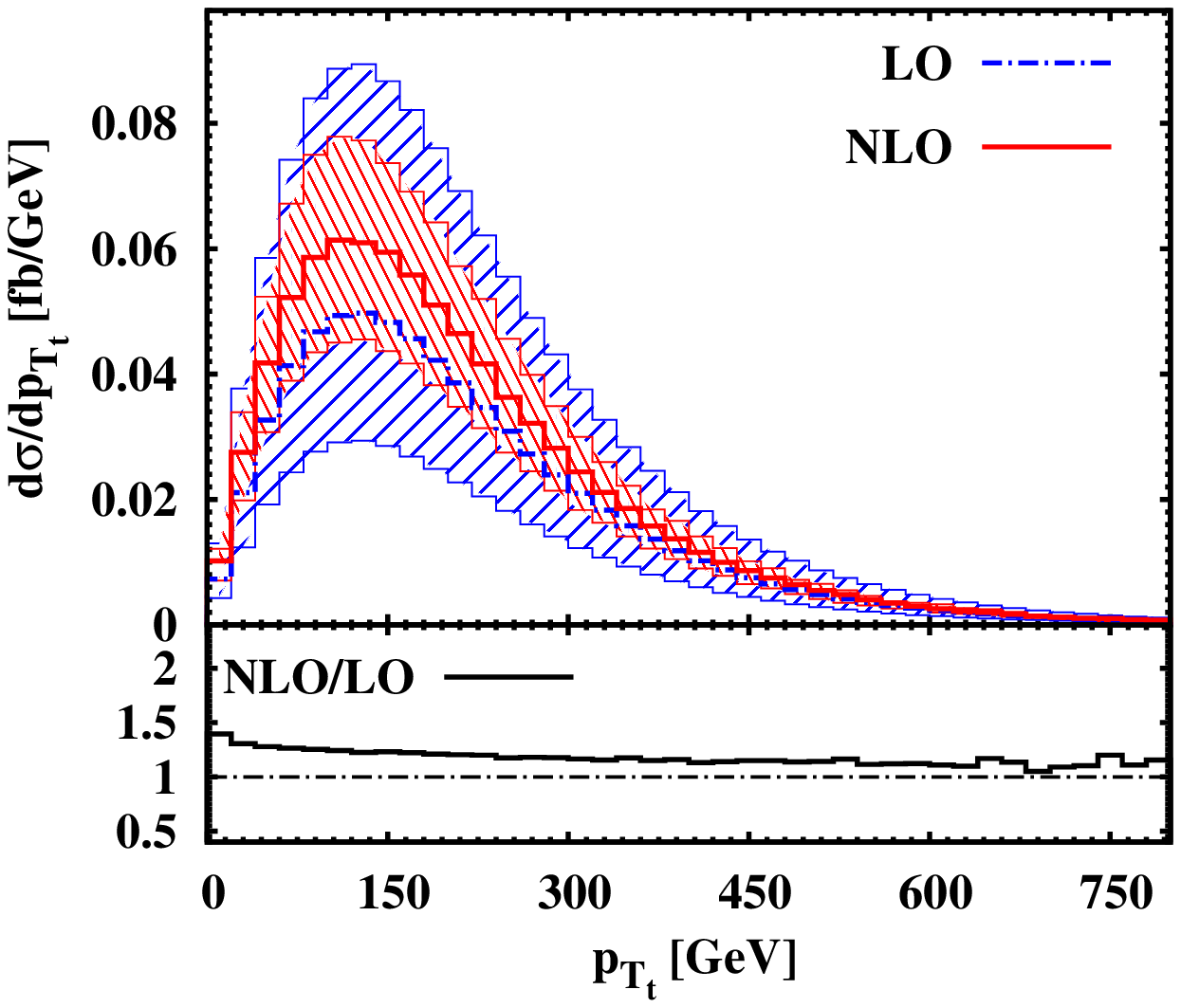}
\includegraphics[width=0.49\textwidth]{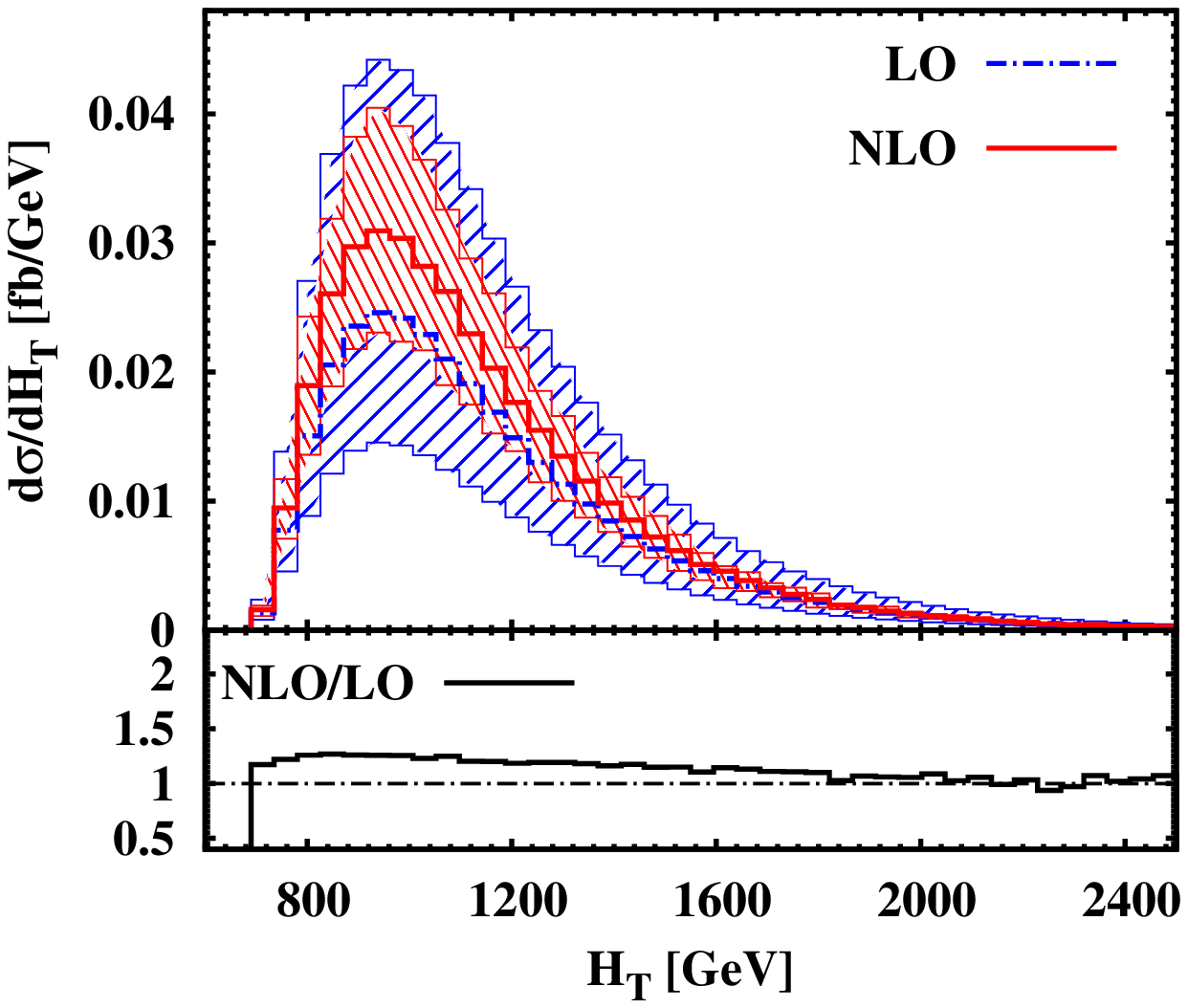}
\end{center}
\caption{\it  \label{lhc:distributions-dynamic}  Averaged differential cross
  section distributions as a function of transverse momentum of the $t\bar{t}$
  pair (upper-left panel)  and the top quark (upper-right panel) for  $ pp \to
  t \bar{t}  t \bar{t} + X$ production at the LHC with $\sqrt{s}= 14
  ~\textnormal{TeV}$.    Also shown is the differential  cross section
  distribution as a function of the total transverse energy of the system
  (lower panel).  The dash-dotted (blue) curve corresponds to the LO, whereas
  the solid (red) one to the NLO result. The scale choice is $\mu_F = \mu_R
  =\mu_0= H_T/4$. The uncertainty bands depict
  scale variation. The lower panels display the differential $\cal K$ factor.}
\end{figure*}
%

\subsection{Differential cross sections for $\mu_0 = H_T/4$}

%
\begin{figure*}
\includegraphics[width=0.49\textwidth]{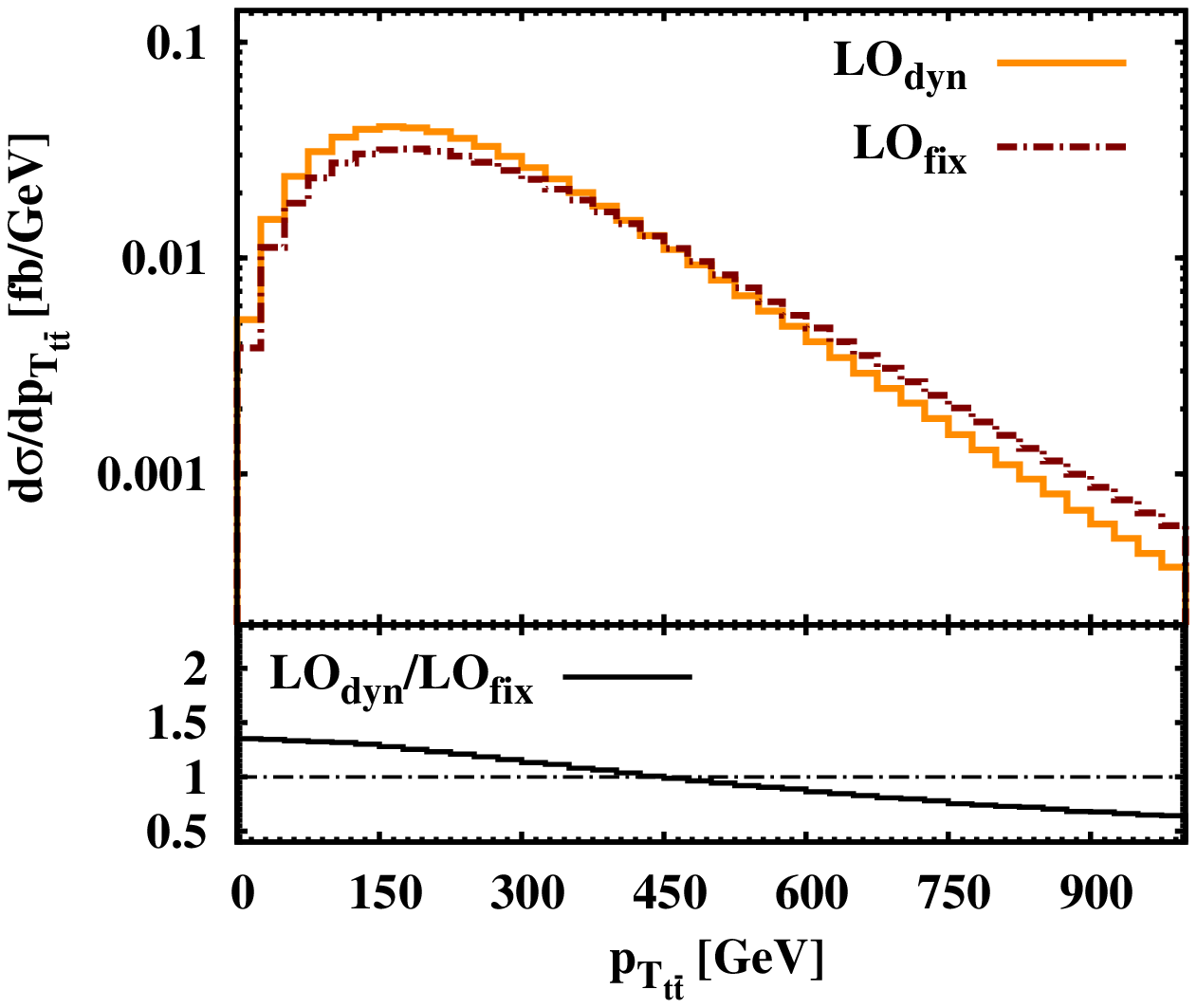}
\includegraphics[width=0.49\textwidth]{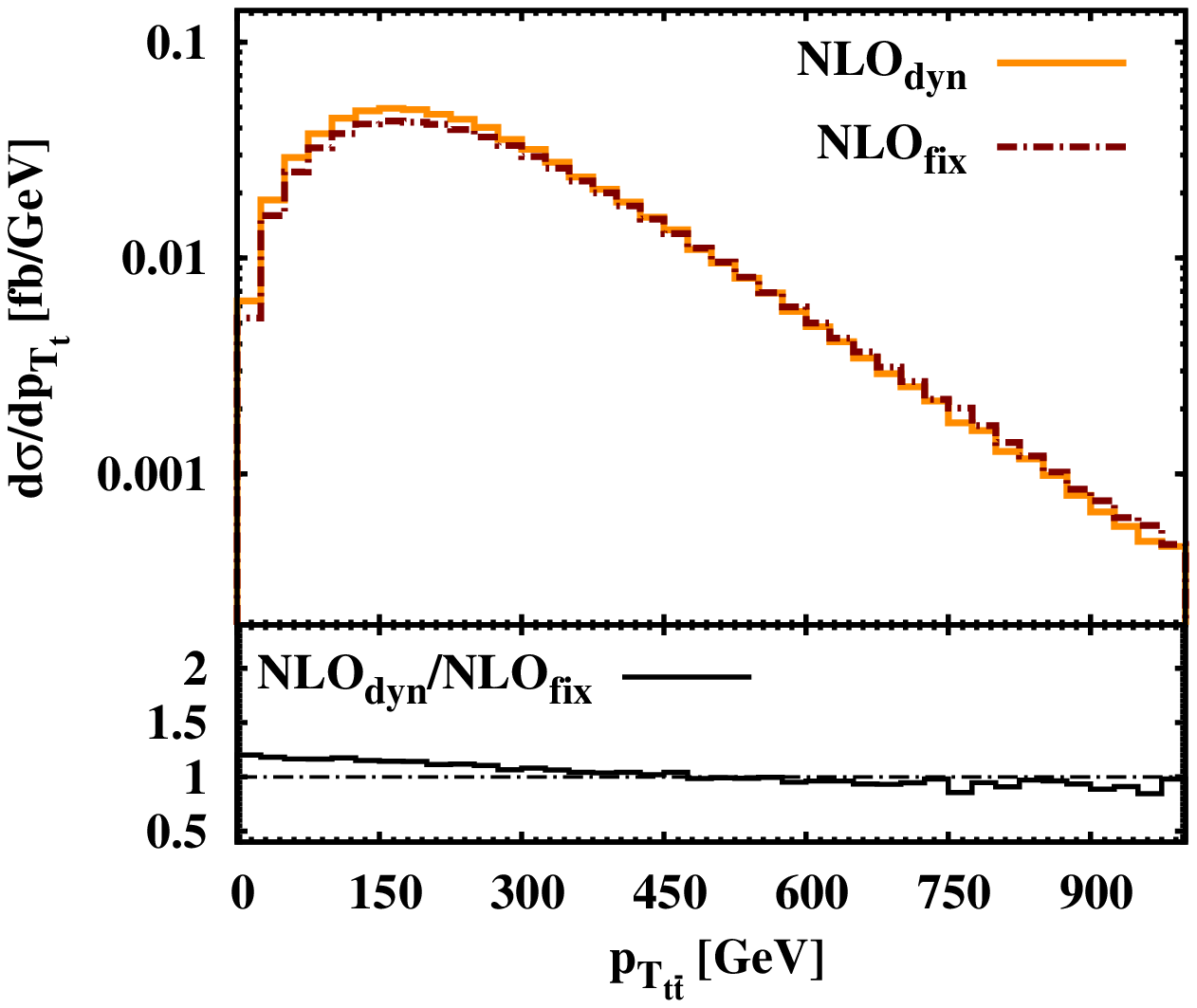}
\caption{\it  \label{lhc:ratio1} Averaged differential cross section
  distributions as a function of transverse momentum of the $t\bar{t}$ pair at
  LO (left panel)  and at NLO (right panel) for  $ pp \to t \bar{t}  t \bar{t}
  + X$ production at the LHC with $\sqrt{s}= 14 ~\textnormal{TeV}$.  The
  dash-dotted (orange) curve corresponds to $\mu_0=2m_t$, whereas the dashed
  (brown) one to $\mu_0=H_T/4$.  The lower panels display the ratio of the
  result with the dynamic scale versus the fixed scale.}
\end{figure*}
\begin{figure*}
\includegraphics[width=0.49\textwidth]{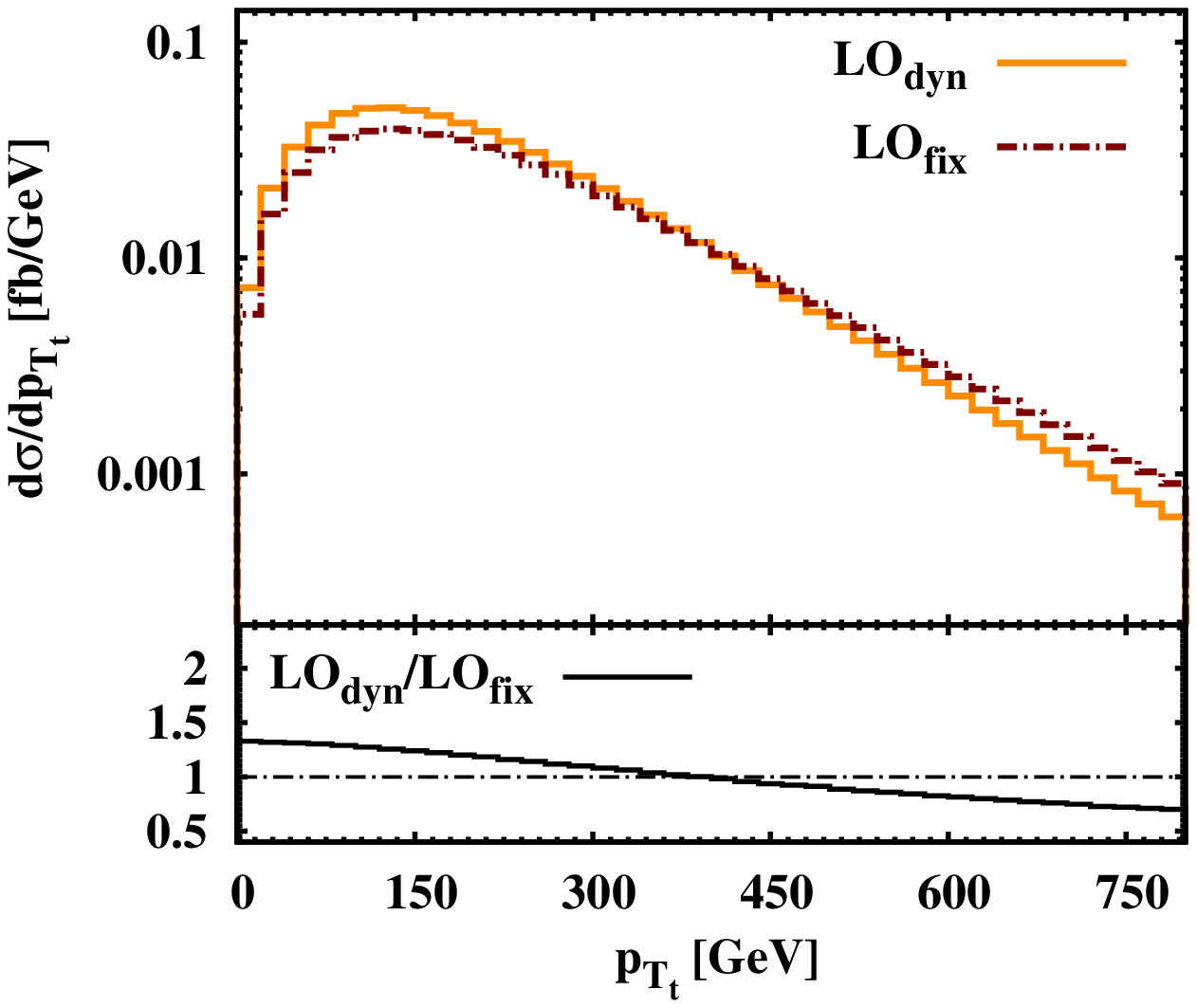}
\includegraphics[width=0.49\textwidth]{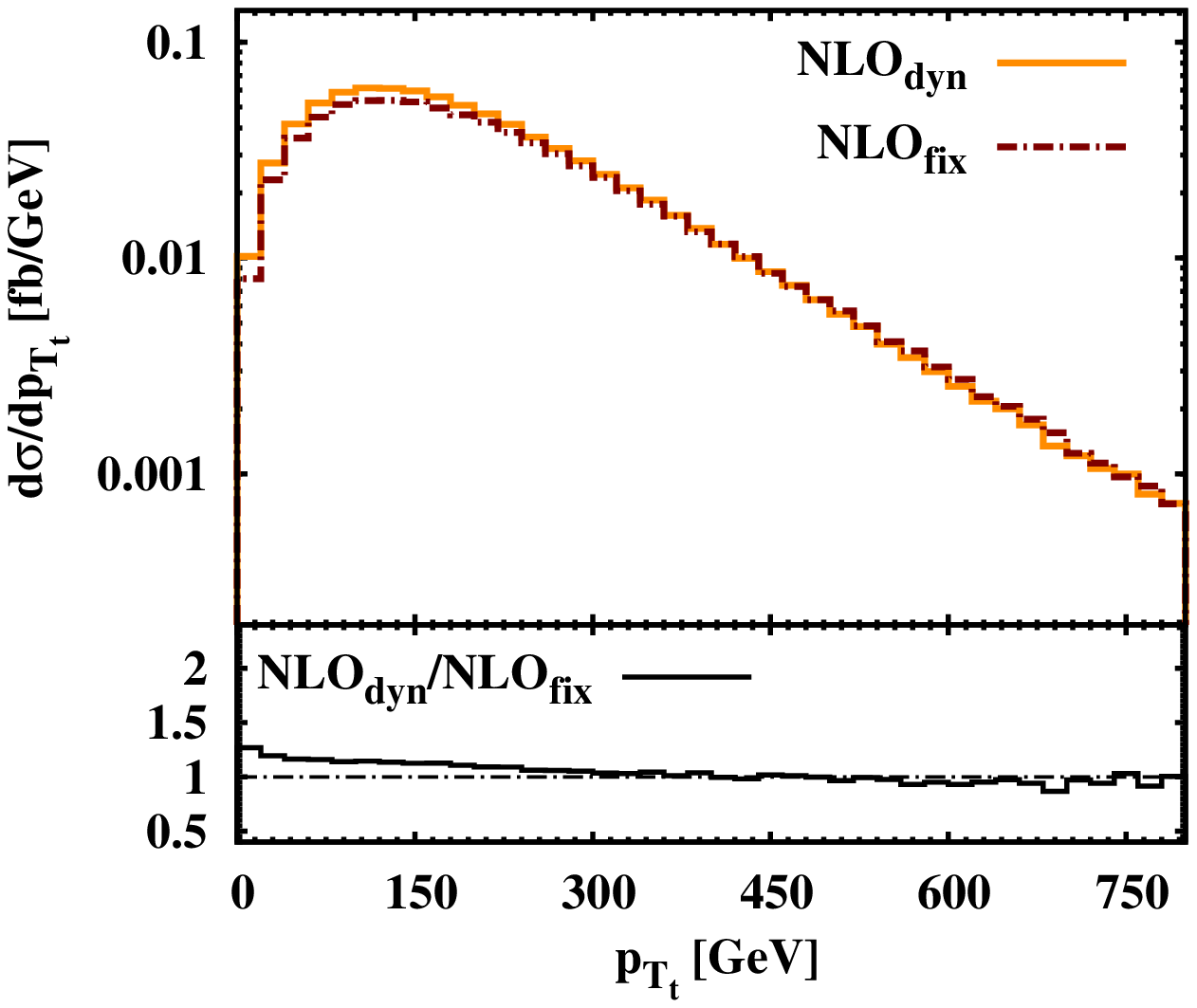}
\caption{\it  \label{lhc:ratio2} Averaged differential cross section
  distributions as a function of transverse momentum of the top quark at LO
  (left panel)  and at NLO (right panel) for  $ pp \to t \bar{t}  t \bar{t} + X$
  production at the LHC with $\sqrt{s}= 14 ~\textnormal{TeV}$.  The
  dash-dotted (orange) curve corresponds to $\mu_0=2m_t$, whereas the dashed
  (brown) one to $\mu_0=H_T/4$.  The lower panels display the ratio of the
  result with the dynamic scale versus the fixed scale.}
\end{figure*}
\begin{figure*}
\includegraphics[width=0.49\textwidth]{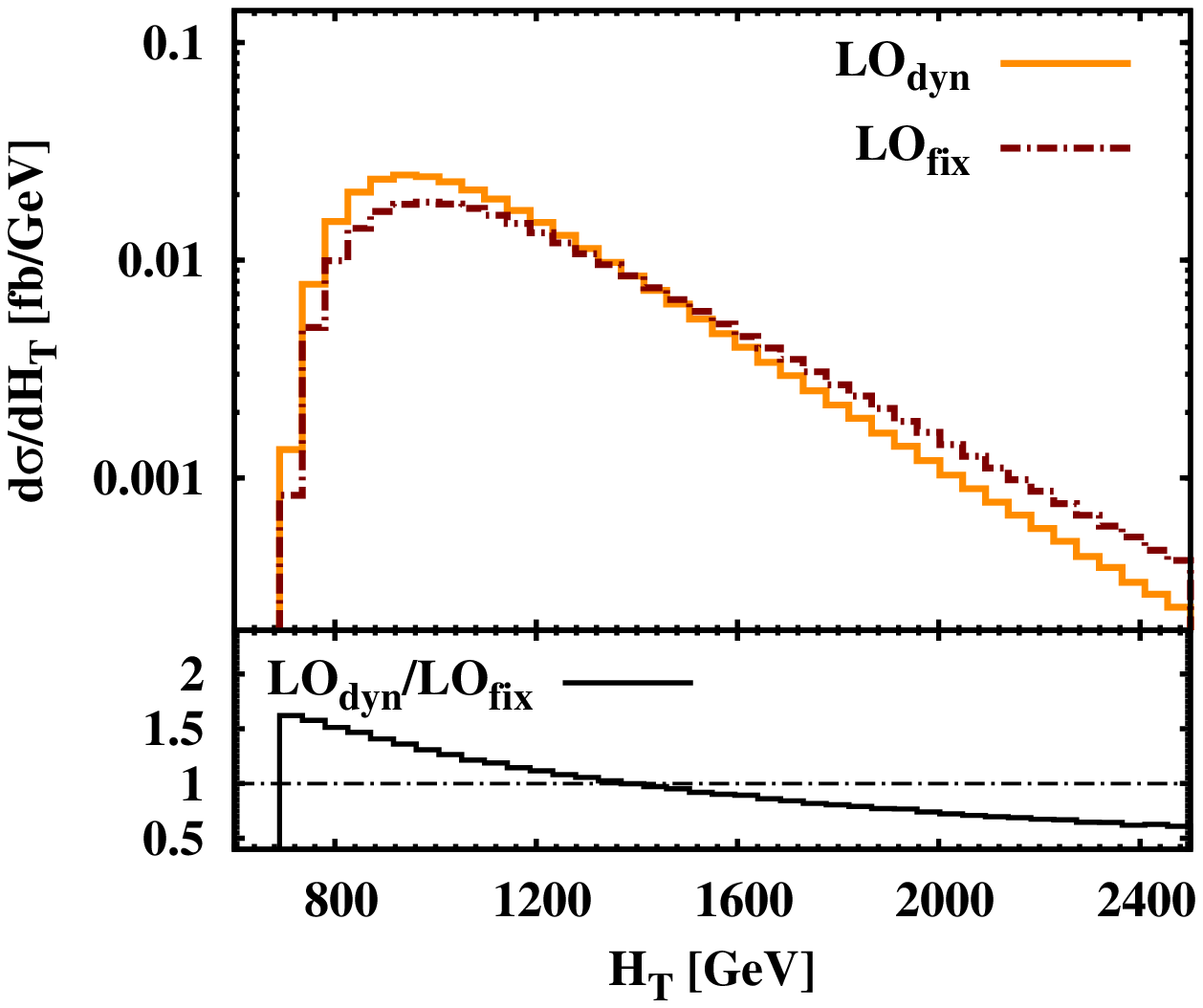}
\includegraphics[width=0.49\textwidth]{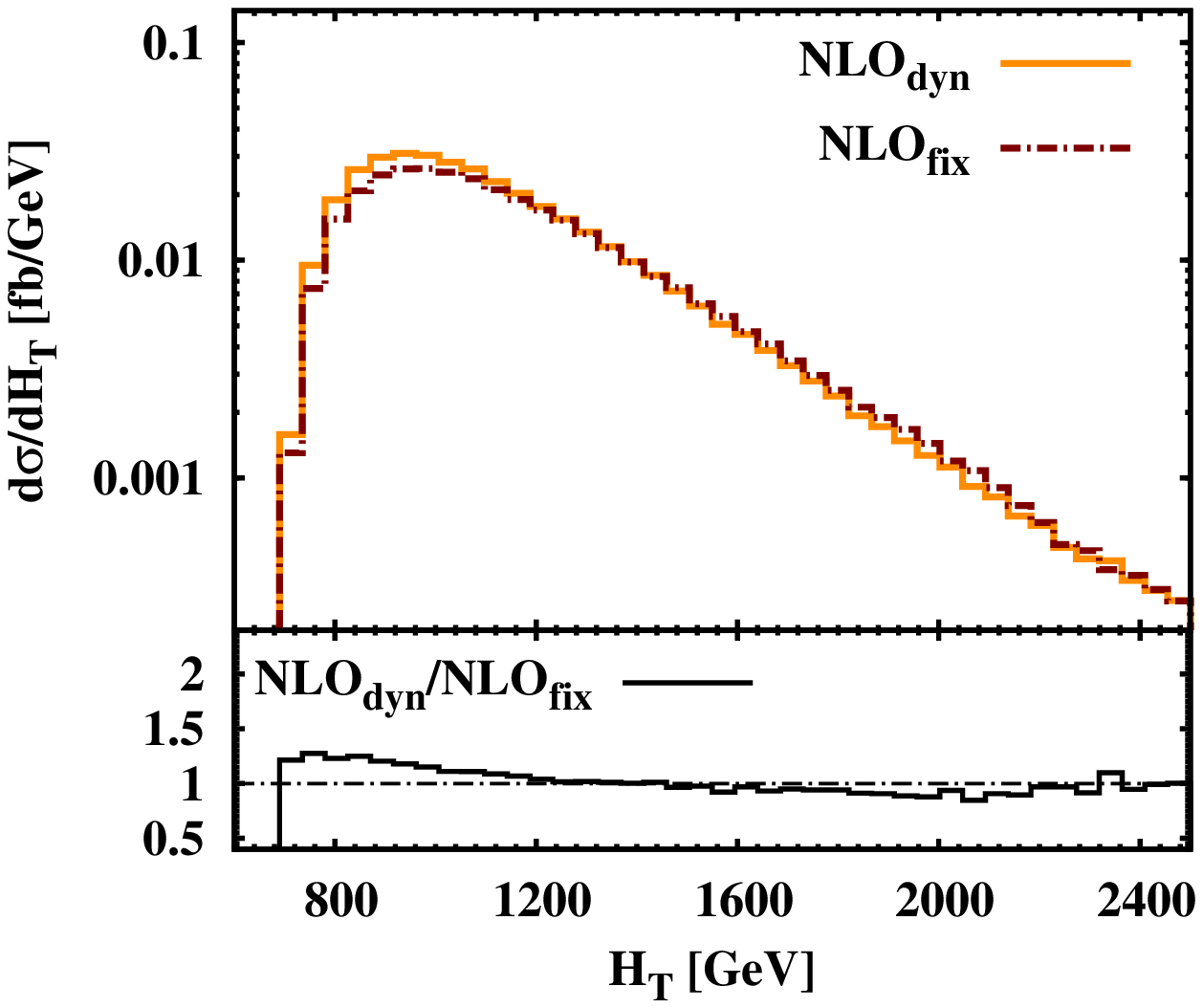}
\caption{\it  \label{lhc:ratio3} Differential cross section distributions as a
  function of total transverse energy of the  $t\bar{t}t\bar{t}$ system at LO
  (left panel)  and at NLO (right panel) for  $ pp \to t \bar{t}  t \bar{t} +
  X$ production at the LHC with $\sqrt{s}= 14 ~\textnormal{TeV}$.  The
  dash-dotted (orange) curve corresponds to $\mu_0=2m_t$, whereas the dashed
  (brown) one to $\mu_0=H_T/4$.  The lower panels display the ratio of the
  result with the dynamic scale versus the fixed scale.}
\end{figure*}
\begin{figure*}
\includegraphics[width=0.48\textwidth]{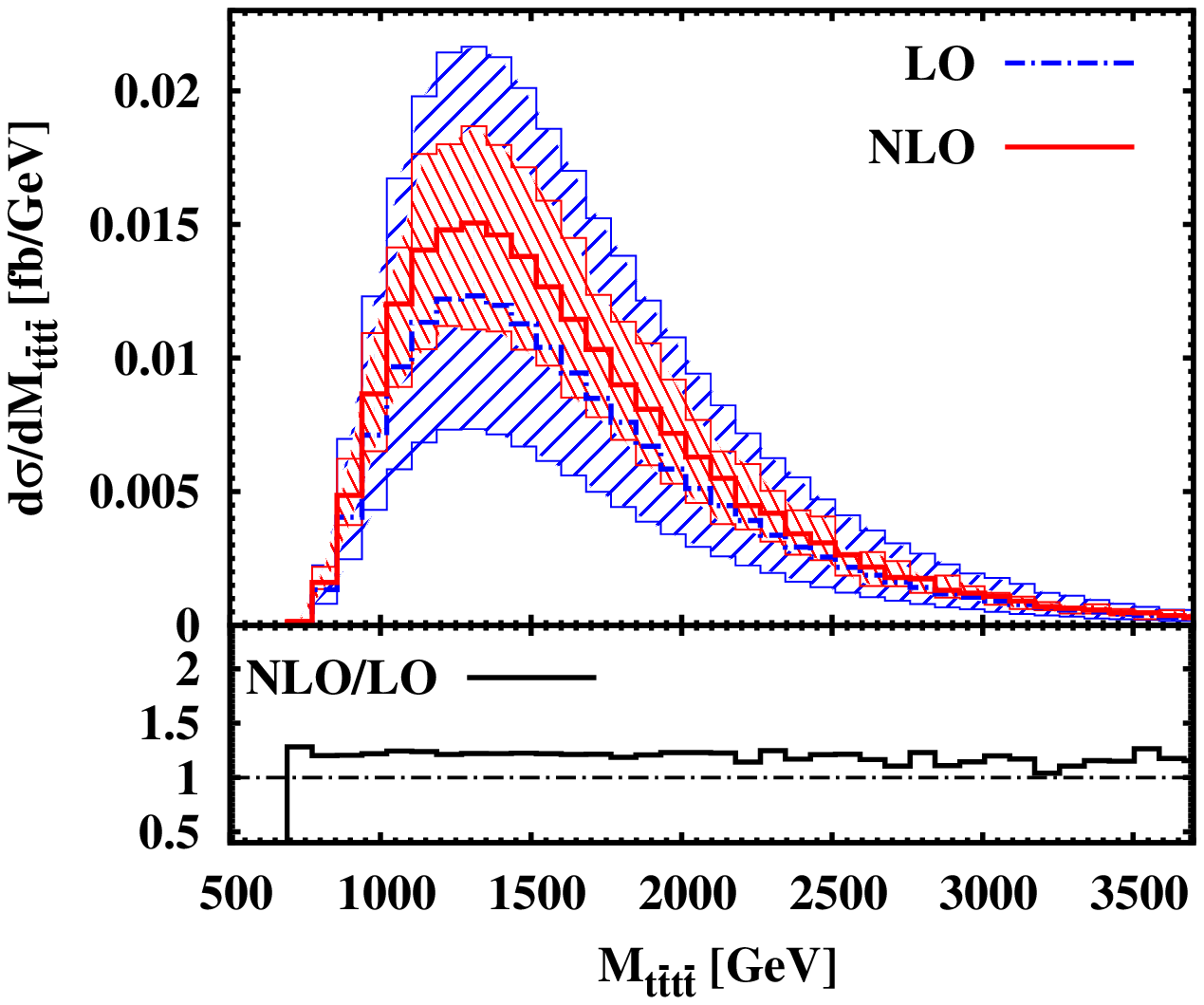}
\includegraphics[width=0.48\textwidth]{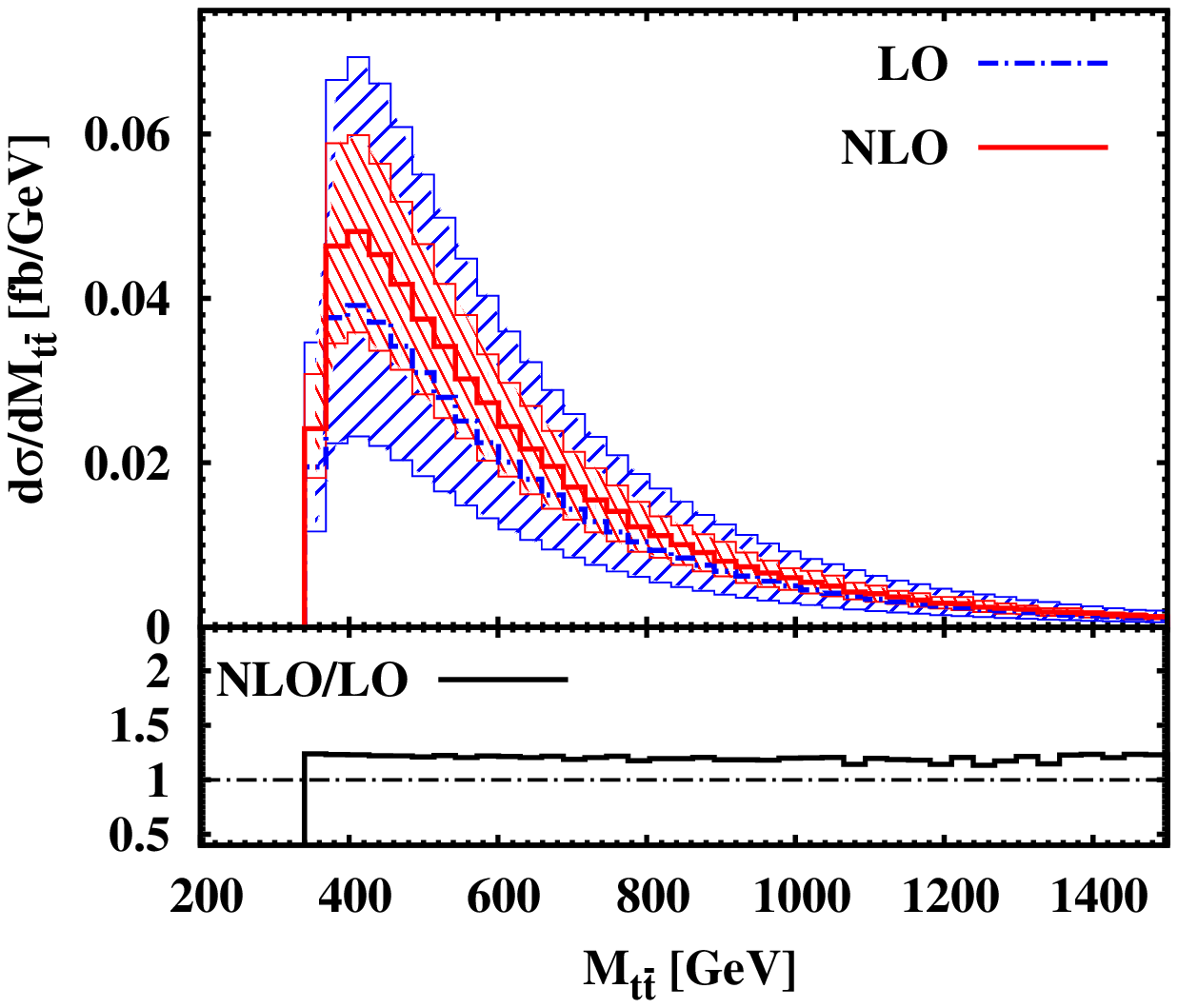}
\caption{\it  \label{lhc:invariant-mass}  Differential cross section
  distributions as a function of the invariant mass of the $t\bar{t}t\bar{t}$
  system (left panel) and the  $t\bar{t}$ pair (right panel) for  $ pp \to t
  \bar{t}  t \bar{t} + X$ production at the LHC with $\sqrt{s}= 14
  ~\textnormal{TeV}$.  The dash-dotted (blue) curve corresponds to the LO,
  whereas the solid (red) one to the NLO result. The scale choice is $\mu_F =
  \mu_R = H_T/4$. The uncertainty bands depict scale variation. The lower
  panels display the differential $\cal K$ factor.}
\end{figure*}
\begin{figure*}
\includegraphics[width=0.49\textwidth]{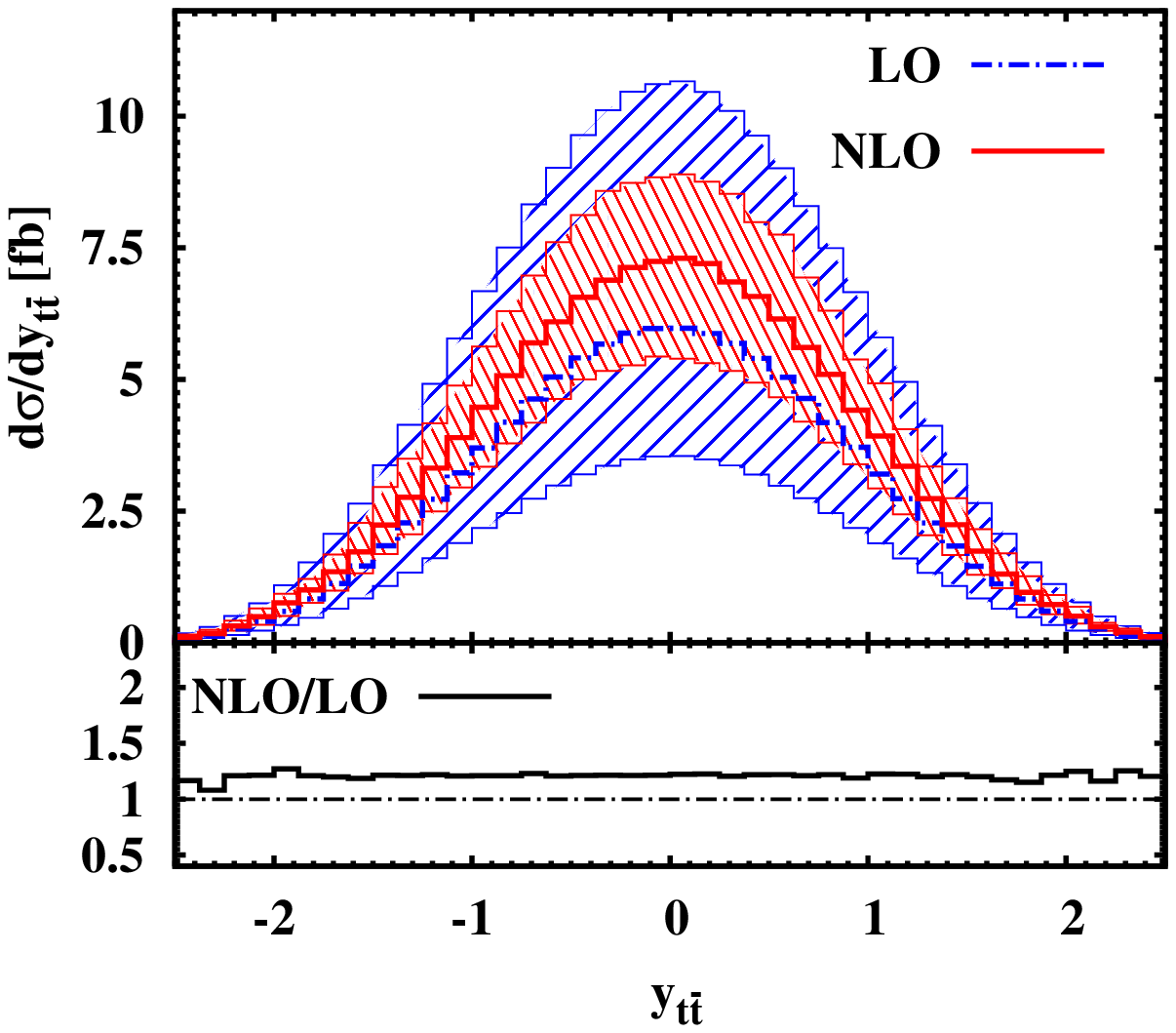}
\includegraphics[width=0.49\textwidth]{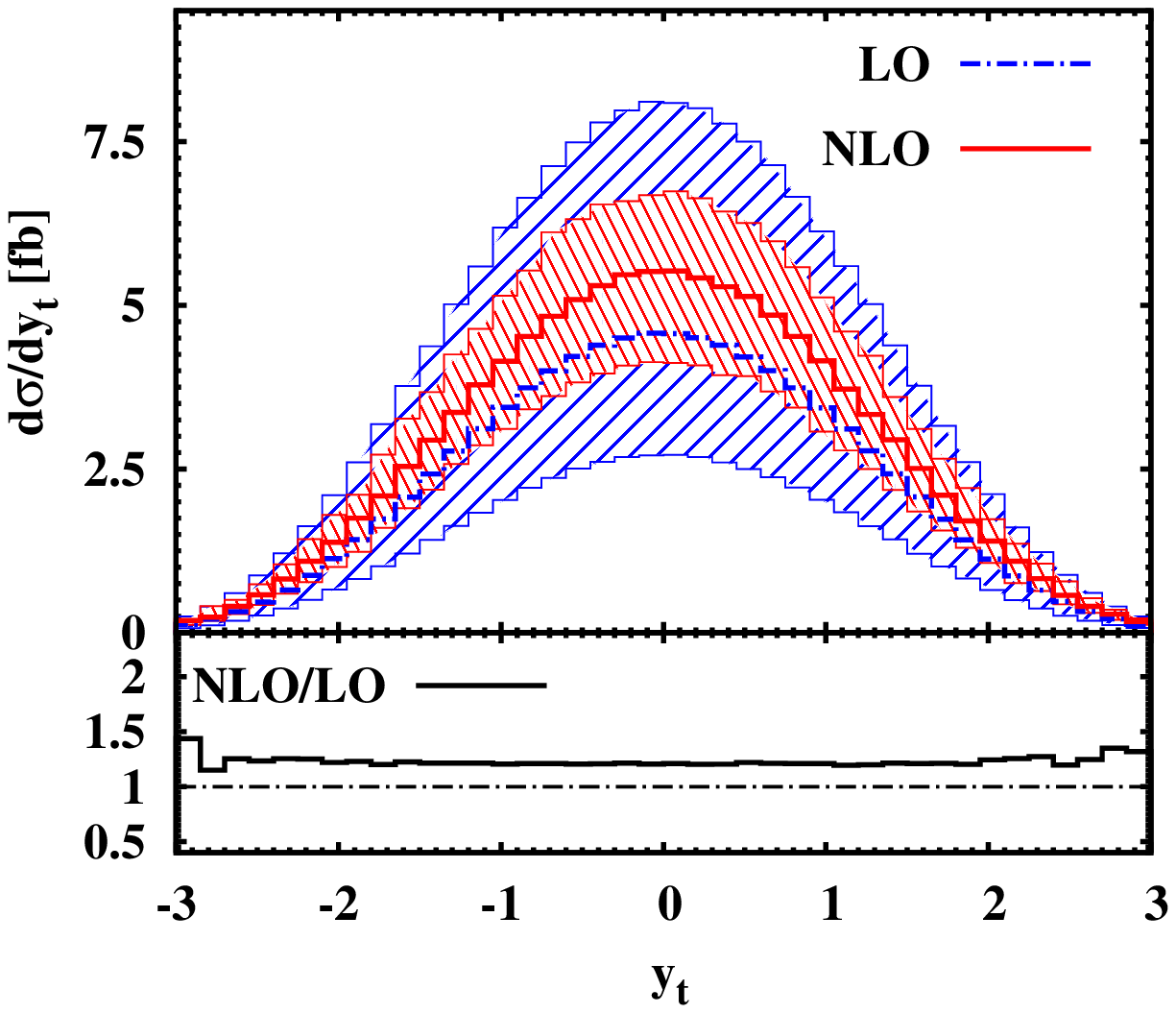}
\caption{\it  \label{lhc:rapidity} Averaged differential cross section
  distributions as a function of rapidity of the  $t\bar{t}$ pair
  (left panel)  and the top quark (right panel) for  $ pp \to t
  \bar{t}  t \bar{t} + X$ production at the LHC with $\sqrt{s}= 14
  ~\textnormal{TeV}$.  The dash-dotted (blue) curve corresponds to the
  LO, whereas the solid (red) one to the NLO result. The scale choice
  is $\mu_F = \mu_R =\mu_0= H_T/4$. The uncertainty bands depict
  scale variation. The lower  panels display the
  differential $\cal K$ factor.}
\end{figure*}
\begin{figure*}
\begin{center}
\includegraphics[width=0.49\textwidth]{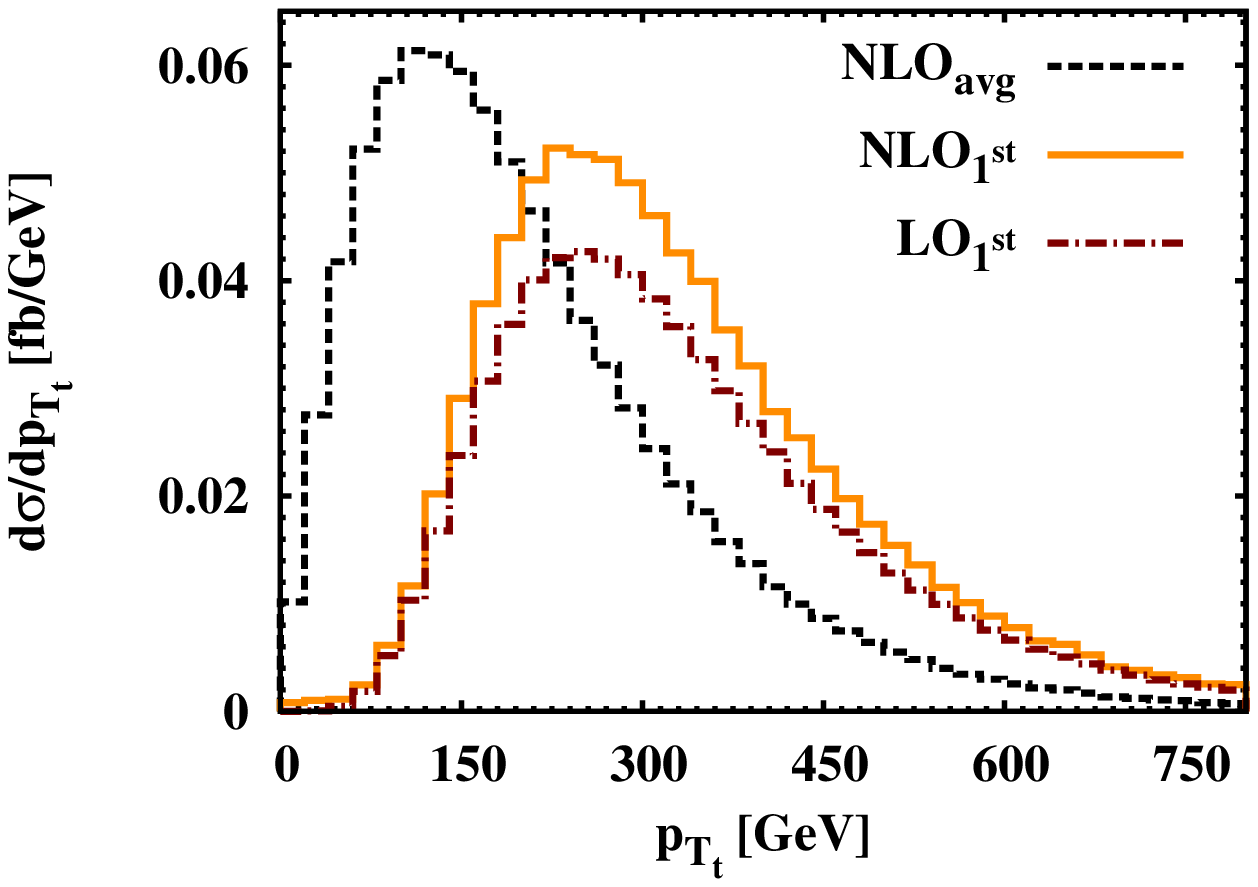}
\includegraphics[width=0.49\textwidth]{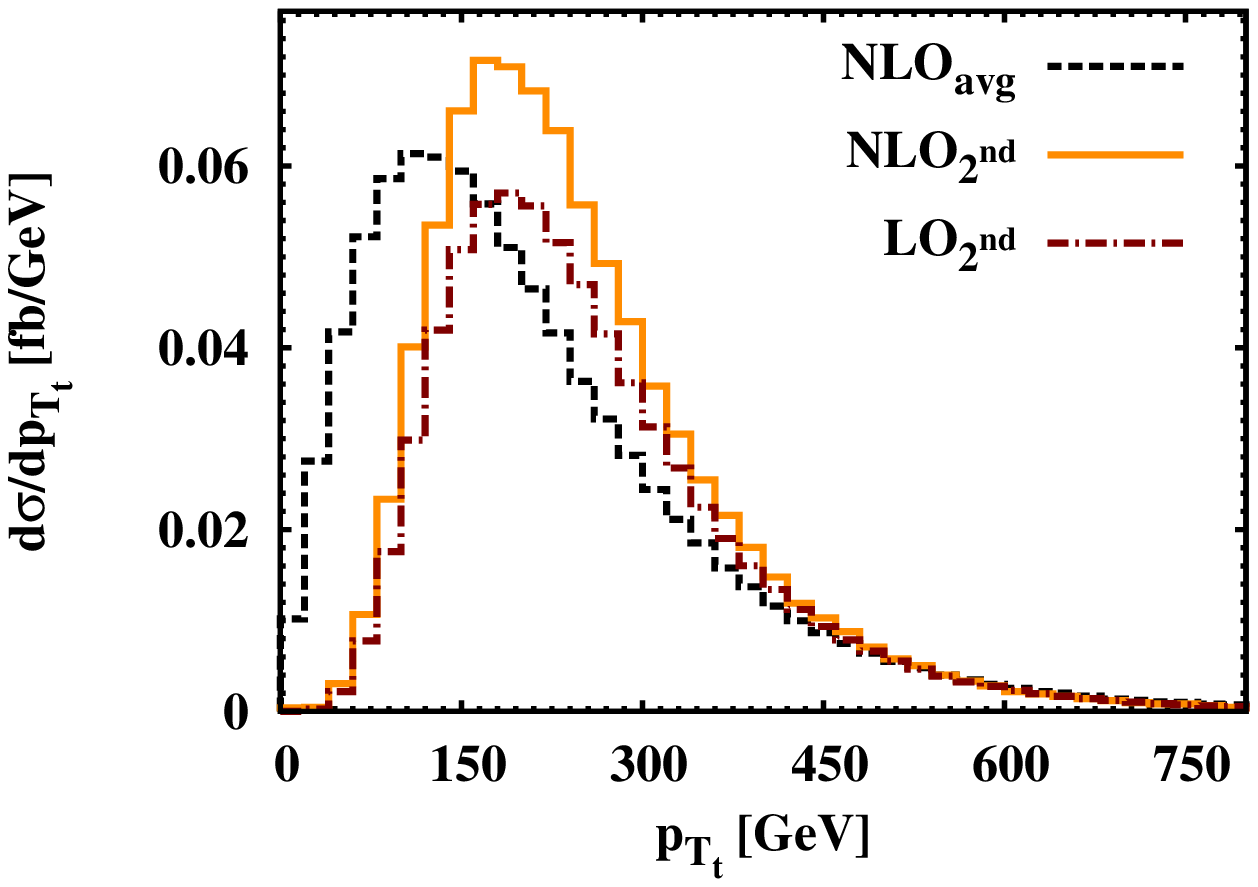}
\includegraphics[width=0.49\textwidth]{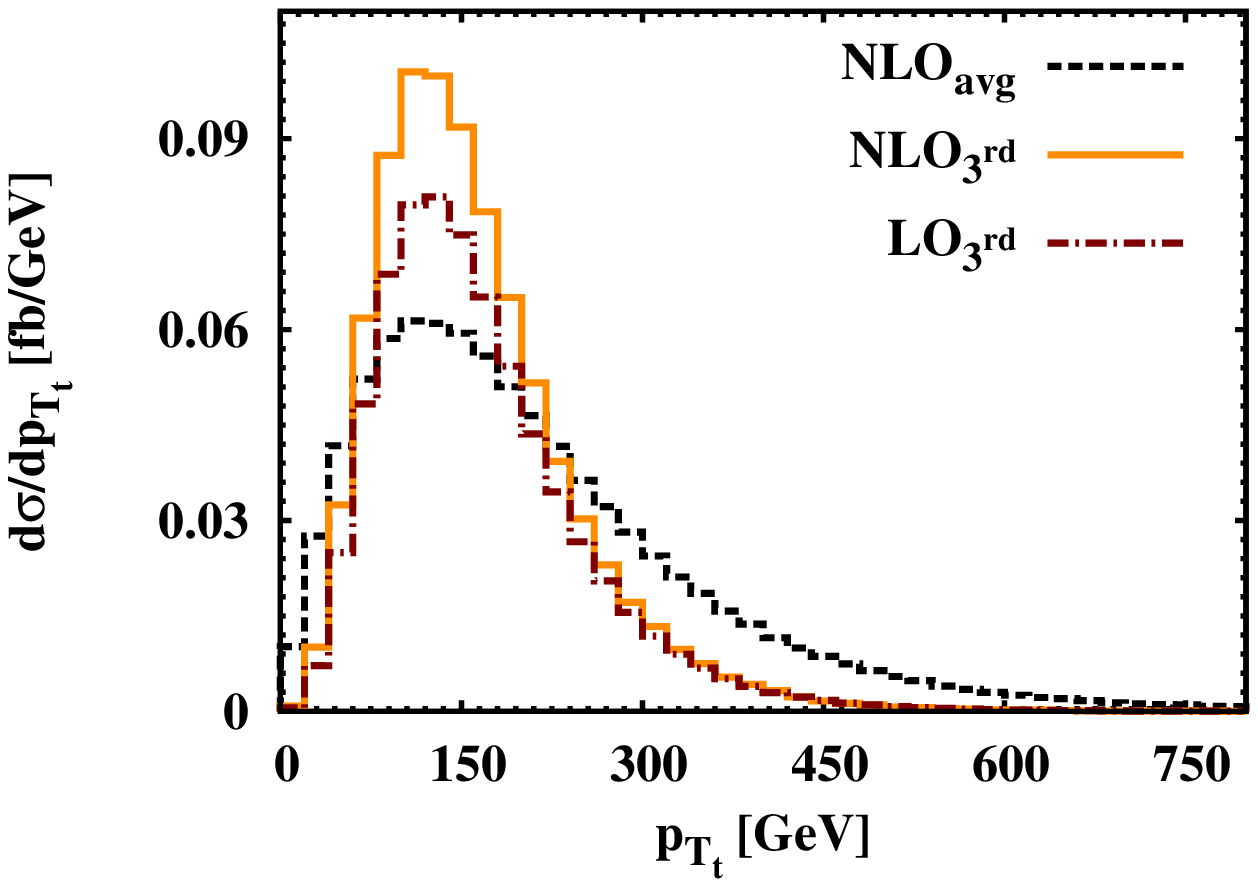}
\includegraphics[width=0.49\textwidth]{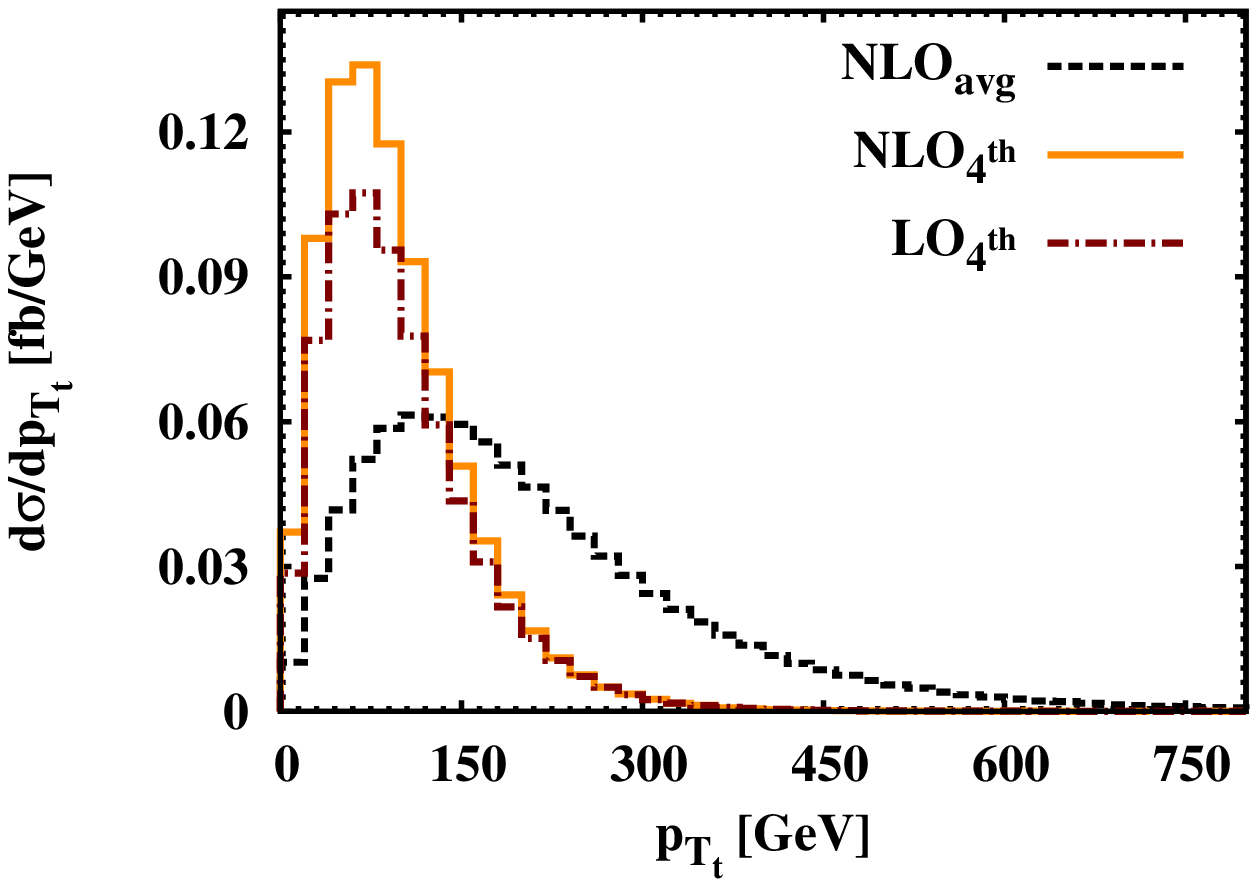}
\end{center}
\caption{\it  \label{lhc:top-pt} Differential cross section
  distributions as a function of the transverse momentum  of the 1st,
  2nd, 3rd and the 4th  hardest top quark at  LO and NLO for  $ pp \to t
  \bar{t}  t \bar{t} + X$ production at the LHC with $\sqrt{s}= 14
  ~\textnormal{TeV}$.  The scale choice is $\mu_F = \mu_R = H_T/4$.
  Also shown is the averaged transverse momentum of the top quark at NLO.}
\end{figure*}
\begin{figure*}
\includegraphics[width=0.49\textwidth]{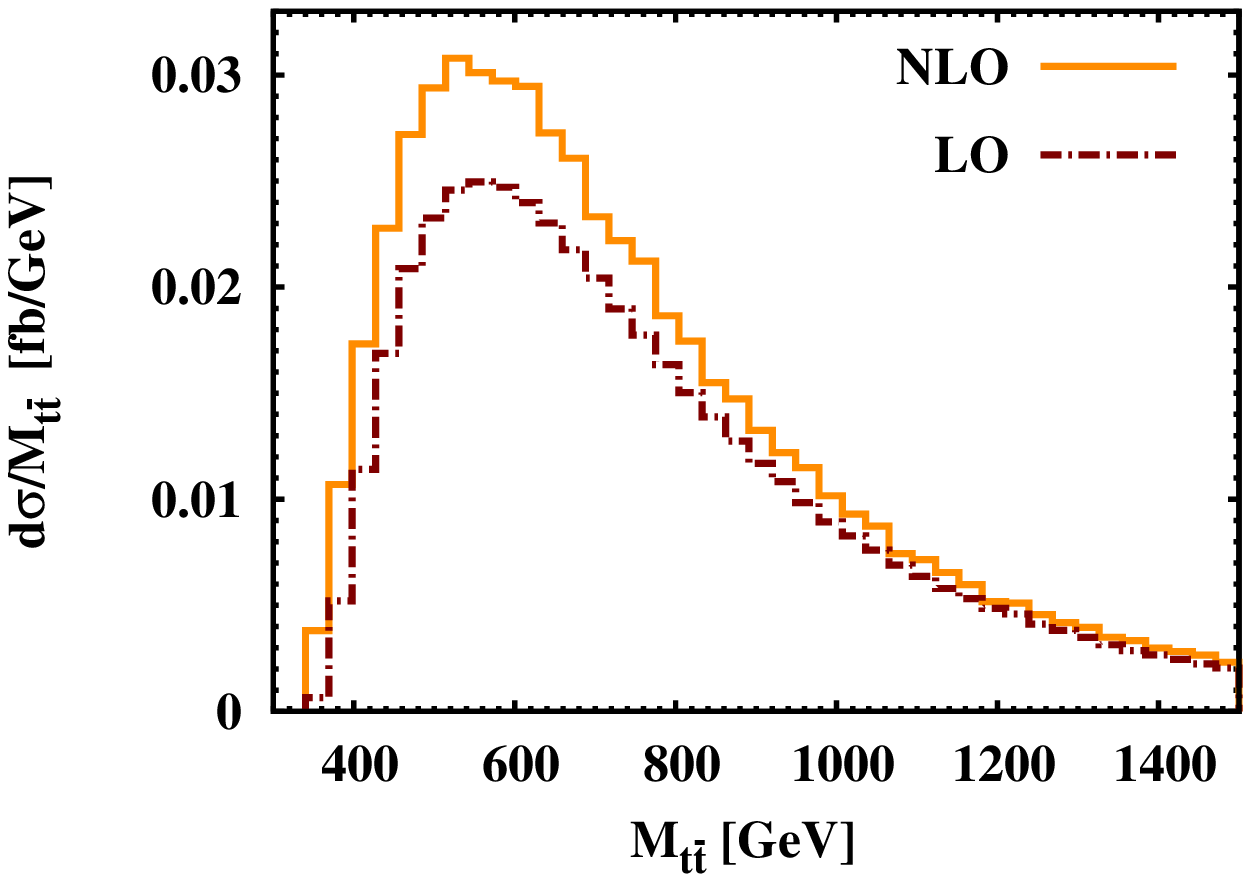}
\includegraphics[width=0.49\textwidth]{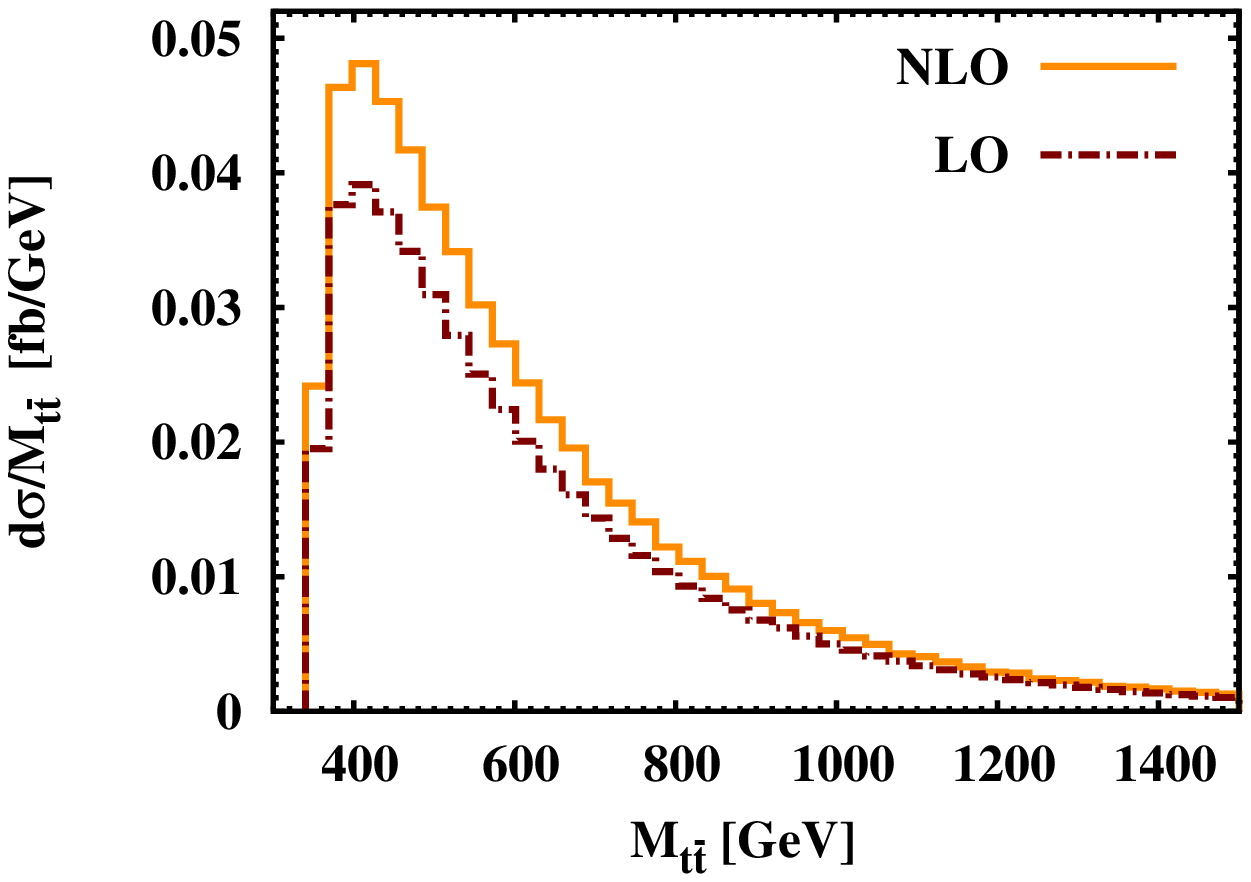}
\caption{\it  \label{lhc:top-inv-max}  Differential cross section
  distributions as a function of the invariant mass of the $t\bar{t}$ pair at
  LO and NLO for $\sqrt{s}=14$ TeV. Left panel: the invariant mass of the two
  top quarks with the highest $p_T$. Right panel: the averaged invariant mass
  of the  $t\bar{t}$ pair. The scale choice is $\mu_F = \mu_R = H_T/4$.}
\end{figure*}
%
As desired, the results for the integrated cross sections  have only
slightly changed in comparison with the fixed-scale case. In the
following, we study the impact of the different scale choice on the
differential cross sections. In Figure \ref{lhc:distributions-dynamic}
the averaged differential distribution as a function of the transverse
momentum of the $t\bar{t}$ pair and the top quark together with total
transverse energy are shown once again, however, this time with the
renormalization and factorization scales calculated on an
event-by-event basis. Instead of $60\%$ distortions that we have
obtained in the previous case for the $p_{T_{t\bar{t}}}$ and $p_{T_{t}}$
distributions, the moderate and positive corrections  of the order or
$20\%$ have been established over the whole range of $p_T$. An
improvement in the differential ${\cal K}-$factor is clearly visible
in both cases. The same conclusions can be drawn for  the $H_T$
distribution. In this case the improvement is even more
impressive. In addition, one can observe that the NLO error bands as
calculated through scale variation nicely fit within the LO error
bands. 

In general, the idea behind a dynamical scale is to accommodate for
multiscale kinematics. In fact, we are not trying to reduce the
${\cal K}$-factor as defined by the ratio of total cross sections, but
rather to obtain a constant one for the distributions. The fixed order
approximation is meaningful, when the improved scale choice affects
NLO cross sections to a much lower extent than the LO ones. We also
demonstrate that with a proper scale, we can already obtain good
results with a constant rescaling of LO distributions. There always
remains the question of how these ideas work in practice. A convincing
example is illustrated on Figure \ref{lhc:ratio1} - Figure
\ref{lhc:ratio3}, for three different observables, where fixed scale
${\cal K}$-factors were most unsatisfactory.  With the luxury of an
NLO cross section, the improvement obtained with a dynamic scale,
which can only be guessed at for the lack of a NNLO result, is
moderate. The change in the shape of LO distributions is, on the other
hand, rather strong. 

We have compared these two scale choices for many other observables and in all
cases the new, dynamic one has decreased the shape difference of distributions
while going from LO to NLO.  As an example, the invariant mass distribution  of
the $t\bar{t}t\bar{t}$ systems and the averaged  invariant mass distribution
of the $t\bar{t}$ pair are illustrated in Figure \ref{lhc:invariant-mass}.
Even though the invariant mass distributions  peak strongly at the
corresponding thresholds, there are non-negligible tails that are extended to
very high invariant mass values. Also for these kinematical regions an almost
flat differential ${\cal K}-$factor has been obtained. In addition, in Figure
\ref{lhc:rapidity}, the angular differential distributions are presented. We
show in particular  the averaged rapidity distribution of the $t\bar{t}$ pair
and the averaged rapidity distribution of the top quark. As one can see from
Figure \ref{lhc:rapidity}, the $t\bar{t}$ pairs and the top quarks are
predominantly produced in the central region.  Also here, the  differential
${\cal K}$-factor is near constant  within the whole range of $y$.

Through the implementation of the dynamical scale large discrepancies between
shapes of distributions at NLO and LO have disappeared.  New  differential
${\cal K-}$factors suggest that the proper scale choice in LO calculations,
which describes the kinematics of the whole process on an event-by-even basis,
together with a suitably chosen global  ${\cal K-}$factor would be sufficient
for this process. This is  good news taking into account that the LO
calculations  are not only less costly, but can be easily merged via CKKW/MLM
procedures  \cite{Alwall:2007fs} with parton shower programs to obtain
complete inclusive hadron level events samples that can be directly compared 
with the experimental data.

Once the impact of the NLO QCD corrections to the differential cross
sections has been established, we turn our attention to the properties
of  the top quarks. In Figure \ref{lhc:top-pt}  a comparison of the
transverse momentum spectra of the first, second, third and the forth
hardest top quark in the $pp\to t\bar{t}t\bar{t} + X$ production at LO
and NLO is displayed. In each plot, the  averaged transverse momentum
of the top quark at NLO is also shown as a benchmark.  As expected,
the shape of the distributions and their peak change when moving from
the  hardest to the softest  top quark configurations.

Finally, the invariant mass of the two top quarks with the highest
$p_T$ is given in Figure \ref{lhc:top-inv-max}.  This observable is
particular interesting in view of new physics searches where the mass
of a new heavy resonance that decays into the top-anti-top pair is
reconstructed as the invariant mass of the two objects with the
highest $p_T$ in the event.  For comparison,  the averaged invariant
mass of the $t\bar{t}$ pair is also presented.
%

\section{Summary and Conclusions}


In this paper we have presented a computation of the NLO QCD corrections to
four top quark production at the LHC.  The total cross section and its scale
dependence have been evaluated for two different scale choices, {\it i.e.} for
the fixed scale $\mu_R=\mu_F=\mu_0=2m_t$ and for the dynamical scale
$\mu_R=\mu_F=\mu_0=H_T/4$. The impact of the NLO QCD corrections on the
integrated cross sections is moderate, of the order of $27\%$ for
$\mu_0=2m_t$ and $21\%$ for $\mu_0=H_T/4$. As to the theoretical uncertainty
of our calculation, the contribution related to unknown higher-order
corrections, as obtained by studying the scale dependence of our NLO
predictions, is of the order of $25\%$. We have also analyzed the theoretical
error arising from different parametrizations of PDFs, being able to quantify
it at the level of $5\%-6\%$, thus well below the uncertainty associated with
scale dependence.

Looking only at the total cross section, which is mostly influenced by final
state production relatively close to the threshold, both scale choices are in
equally good shape and the results agree well within the corresponding
theoretical errors. On the other hand, differential cross sections show large
differences in shape, with distortions up to $80\%$ observed within our
fixed-scale setting. In particular, large negative corrections are clearly
visible in the tails of several distributions. Thus, an accurate description
of the shapes of observables can be given only via full NLO QCD computation
in this case. Instead, adopting our dynamical scale choice, results 
have  moderate, positive and almost constant corrections of
the order of $20\%$ for all the investigated observables. This fact suggests
that the proposed dynamical scale efficiently accommodates for the multiscale
kinematics of the process. 

Well-behaved as it is, the proposed dynamical scale has a number of advantages
for phenomenological studies. Indeed, it can be used within a LO calculation,
together with a suitably chosen global ${\cal K-}$factor, to obtain results
that well approximate the full NLO QCD calculation and can be merged with
parton shower programs to obtain realistic hadronic events, directly
comparable with the experimental data for new physics searches. Particularly
interesting observables in this sense are the invariant mass and the total
transverse energy of the $t\bar{t}t\bar{t}$ system, together with the
invariant mass of the two hardest top quarks.

As a final remark we point out that, despite its relatively small cross
section, a good theoretical control over the SM $pp \to t\bar{t}t\bar{t}$
background can be phenomenologically relevant. Our NLO QCD predictions are of
the order of $17 \pm 4 \,[{\rm scales}] \pm 1 \, [{\rm PDF}]$ fb for $\sqrt{s}
= 14$ TeV. For comparison, typical predictions of new physics scenarios such
as effective four-top interactions, Kaluza-Klein gluons or the so-called
top-philic $Z^\prime$ ({\it i.e.}, a $Z^\prime$ which couples to third-family
quarks only) are set in the range $5-100$ fb for $m_{new} = 1$ TeV and $1-20$
fb for $m_{new} = 1.5$ TeV, where $m_{new}$ is the mass of the new heavy
particle or, more generally, the energy scale associated with new physics
effects. For  masses greater than $2-3$ TeV rates are below 1 fb
\cite{Brooijmans:2010tn}. We believe that a NLO analysis of  $pp \to
t\bar{t}t\bar{t}$ at the LHC is a necessary step towards a correct
interpretation of the possible signals of new physics that may arise in this
channel. 

%

\acknowledgments

%
The calculations have been performed on the Grid Cluster of the Bergische
Universit\"at Wuppertal, financed by the Helmholtz - Alliance Physics at the
Terascale and the BMBF.  

The work of M.W. was supported by the Initiative and Networking Fund of the
Helmholtz Association, contract HA-101 (Physics at the Terascale).

The work of G.B. was supported by the DFG Sonderforschungsbereich/Transregio 9
Computergest\"utzte Theoretische Teilchenphysik.

 We thank Ansgar Denner for  providing us with   partial results for the NLO
 QCD corrections to the $pp \to t\bar{t}b \bar{b} + X$ production with
 $\mu^2_R=\mu^2_F = \mu^2_{0}= m_t \sqrt{p_T(b) \cdot p_T(\bar{b})}$   that
 helped to cross check our implementation of the dynamical scale in  the
 \textsc{Helac-Nlo} framework. 

We also would like to thank Alberto Guffanti and Maria Ubiali for discussions
concerning PDFs.

\providecommand{\href}[2]{#2}\begingroup\raggedright\endgroup

\end{document}